\documentclass[aps,pra,showpacs,twoside,twocolumn,longbibliography,10pt]{revtex4-1}
\usepackage[colorlinks=true, citecolor=red, urlcolor=blue ]{hyperref}
\usepackage{epsfig,newlfont,amssymb,amsfonts,amsmath,bm,subfigure,palatino,mathtools,amsthm,braket,times,soul,enumitem,color}
\usepackage[normalem]{ulem}
\newcommand{\stkout}[1]{\ifmmode\text{\sout{\ensuremath{#1}}}\else\sout{#1}\fi}
\usepackage[english]{babel}
\usepackage[utf8]{inputenc}
\usepackage{array}
\usepackage{xcolor}
\usepackage{graphics}
\newtheorem{theorem}{Theorem}

\begin{document}

\title{Quantum sensing of even- versus odd-body interactions}
\author{Aparajita Bhattacharyya, Debarupa Saha, Ujjwal Sen}
\affiliation{Harish-Chandra Research Institute, A CI of Homi Bhabha National Institute, Chhatnag Road, Jhunsi, Prayagraj 211 019, India}

\begin{abstract}
We analyze the scaling of quantum Fisher information with the number of system particles in the limit of large number of particles, as a function of the number of parties interacting with each other, for encoding Hamiltonians having $k$-body interactions, where $k$ is arbitrary. We find that estimation of coupling strength of such arbitrary-body encoding Hamiltonians provide a super-Heisenberg scaling  that increases monotonically with an increase in the number of interacting particles, in the limit of large number of system particles. Moreover, we also ask if genuine multiparty entanglement is indispensable in attaining the best metrological precision if we employ non-local terms in the Hamiltonian. We identify  a dichotomy in the answer. Specifically, we find that Hamiltonians having odd-body interactions necessarily require genuine multipartite entanglement in probes to attain the best metrological precision, but the situation is  opposite in the case of Hamiltonians with even-body interactions. The optimal probes corresponding to Hamiltonians that contain even-body interaction terms, may be entangled, but certainly not so in all  bipartitions, and particularly, it is possible to attain optimal precision using asymmetric probes. Asymmetry, which therefore is a resource in this scenario rather than genuine multiparty entanglement, refers to the disparity between  states of  local parts of the global system. Thereby we find a complementarity in the requirement of asymmetry and genuine entanglement in optimal probes for estimating strength of odd- and even-body interactions respectively. Additionally, we provide an upper bound on the number of parties up to which one can always obtain an asymmetric product state that gives the best metrological precision for even-body interactions. En route,  we find the  quantum Fisher information in closed form for two- and three-body interactions for  arbitrary number of parties.
We also provide an analysis of the case when the Hamiltonian contains local fields and up to $k$-body interaction terms, where the strength of interaction gradually decreases with an increase in the number of parties interacting with each other. Interestingly, we find a similar dichotomy in the nature of the optimal probe in this case as well, i.e. for encoding Hamiltonians with up to even- and odd-body interactions.
Further, we identify conditions on the local component of the Hamiltonian, for which this dichotomy is still shown to exist for two- and three-body encoding Hamiltonians with arbitrary local dimensions.
\end{abstract}

\maketitle
\section{Introduction}
The goal of quantum metrology~\cite{gen1,gen2,gen3,gen4,gen5,gen6,gen7,gen8,gen9,Maccone1,noi1,gen11,gen12,noi2,noi5,gen10,gen13,noi6,noi8,noi7,noi9,noi10,noi11,ref1,noi3,noi4,h1,ref2} is to estimate an encoded parameter with minimum possible error. There is a lower bound in estimating the error of the encoded parameter, given by the quantum Cram\'er-Rao bound~\cite{Wooters,Braunstein1,holevo,macconerev,review1}. 
The bound is given in terms of a quantity, referred to as the quantum Fisher information (QFI), which can be identified as the amount of information that can be decoded from the process of estimating the relevant parameter.

Our aim is to estimate the coupling strength  of arbitrary-body interacting encoding Hamiltonians. 
Interacting many-body Hamiltonians~\cite{manyb1,manyb2,manyb3,manyb4,manyb5} are crucial both fundamentally and technologically in performing numerous quantum mechanical tasks like quantum error correction~\cite{er1,er2,errora,errorb,er3}, creation of entanglement~\cite{en1,en2}, etc. Therefore it is important to look at the metrological aspect of unitary encoders, in which the Hamiltonians contain non-local terms~\cite{int1,int2,int3,luccapezze1,gen13,1,2,3,4,6,5,7,luccapezze2,8,9,10,int5,int6,int4}. 
It is worth noting that in a previous work~\cite{int1}, the authors provided bounds on the QFI corresponding to a multibody encoding Hamiltonian, where the bound is attainable using a specific genuine multipartite entangled state. This result was derived under the assumption that all eigenvalues of each single-body operator are non-negative. However, there is a striking dissimilarity between their work and ours. In particular, we consider the single-body terms to be Pauli operators, which include negative eigenvalues. This leads to a distinction in the nature of the optimal probe for  odd- and even-body interacting encoding Hamiltonians. Therefore, the results obtained in our work are new and fundamentally different from those in Ref.~\cite{int1}.


The maximum QFI and  features of optimal probes for encoding Hamiltonians having single-body terms, have already been studied in literature~\cite{Maccone1}.  We aim to find the scaling of QFI with the number of particles when the encoding Hamiltonian has two-, three- or higher-body interaction terms, in the limit of large number of system particles.  
We find that the optimal scaling of QFI monotonically increases with the number of particles interacting with each other. Specifically, we find super-Heisenberg scaling of QFI, viz. $\sim N^{2k}$, with $N$ particles and $k$-body interactions for $k>1$. Along with the scaling of QFI for optimal probes, we also provide an analysis of the maximum precision with symmetric product probes for completeness.
The other 
main goal in this paper is to find whether the role of genuine multiparty entanglement prevails for $k$-body interactions, where $k>1$.  Our finding suggests that the answer to this has a dichotomy. In particular, we find that for odd values of $k$, the optimal input probes necessarily possess genuine multiparty entanglement (GME)~\cite{entrev}, but for even $k$, the opposite is true, viz. the optimal probes are not genuinely multiparty entangled. 
Intriguing features are observed in the patterns of the optimum input probe for even values of $k$. These optimal states may possess some entanglement, but always have vanishing GME. Unlike the case of Hamiltonians containing only local terms, where the presence of GME~\cite{ghz1,ghz2} was a necessity in the optimum probe~\cite{ent1,ent2,ent3,ent4,ent5}, here we find that the best precision can be achieved with states having zero GME, and in certain cases even with product states. Further, for each even value of $k$, we find certain ranges of $N$, including the large-$N$ regime, where $N$ denotes the number of system particles, for which the optimum probe is a zero GME asymmetric state, possibly a product state. Asymmetry, rather than genuine multiparty entanglement, is thus a resource in quantum parameter estimation with $k$-body interactions for even $k$.
Interestingly, we find a complementarity in the resource requirement in optimal probes for estimating the strength of even- and odd- body interactions.  Along the way, we also provide the QFI in closed form for two- and three-body interactions, for arbitrary multipartite systems, for the two types of optimum inputs.
We also provide an analysis of the case when the Hamiltonian contains lcal fields and up to $k$-body interaction terms, where the strength of interaction gradually decreases with an increase in the number of parties interacting with each other. Interestingly, here also we find a similar dichotomy in the nature of the optimal probe for up to even- and odd-body interactions.
We first analyzed the scenario where each local party is a qubit and later extended our results for two- and three-body interaction, to include cases when the local dimension is arbitrary. 
\vspace{0.2cm}

\section{Probe Hamiltonians}
A preliminary discussion on parameter estimation protocol and results from previous literature using local encoding Hamiltonians is provided in Appendix~\ref{prelim}.
In this work, we inspect two types of scenarios. Firstly, we consider an $N$-partite system, whose evolution is governed by a Hamiltonian that has solely $k$-body interaction terms where $k$ is arbitrary and can take any integer value from $1$ to $N$. 
We consider another physically relevant situation, where the Hamiltonian  has up to $k$-body interaction terms, where the interaction strength gradually decreases with an increase in the number of parties interacting with each other.
In the first scenario considered, which we denote as case I, the encoding Hamiltonians are of the form $J\sum_{i=1}^{N} \sigma^i_z$, $J\sum_{i,j=1, j>i}^{N} \sigma^i_z \sigma^j_z$, $J\sum_{i,j,l=1, l>j>i}^{N} \sigma^i_z \sigma^j_z \sigma^l_z$, respectively for values of $k$ equal to $1,2,3$, and so on.
For $k=1$, the Hamiltonian essentially consists of local fields, whereas for $k>1$, the Hamiltonians explicitly contain interaction terms. Each local subsystem is considered to be a qubit, and each local Hamiltonian is taken to be a Pauli-$\sigma_z$ operator.  
We denote such $k$-body interacting Hamiltonians comprising of $N$ parties by the notation $h_k^{(N)}$. Our goal is to estimate the parameter, $J$, which represents a  uniform coupling strength among the different parties. For $k=1$, $J$ represents a field strength. In the second situation, which we denote by case II, we use the notation $\widetilde{h}_k^{(N)}$, to represent Hamiltonians with up to $k$-body interaction terms. Explicitly, the Hamiltonian is given by
 $ \widetilde{h}_k^{(N)} = \sum_{i=1}^k x^i h_i^{(N)} $,
where $x^i$ denotes the common coupling strength for all the $i$-body interaction terms, with $0\le x \le 1$. Here we estimate the parameter $x$. In our analysis, we always consider $N\ge k$. Often, $N\gg k$ is assumed. 

Parameter estimation with multibody encoding Hamiltonians is a well-motivated arena of research. Progress in  theoretical, computational and experimental methods in the recent years have enabled to go beyond two-body forces to efficiently probe three- or higher-body interactions, which can be typically realized in low dimensions~\cite{ldim1,ldim2}, atomic and nuclear systems~\cite{3bodyrev}, or even artificially engineered in cold atom systems~\cite{3bodyrev}. Effective $k$-body interactions, where $k$ is arbitrary, arise when high-energy degrees of freedom are integrated out, or when the underlying two-body potential is replaced by an effective pseudopotential~\cite{multibody00,multibody01}.  The induced interactions play a crucial role across a wide range of physical contexts, from nuclear and high-energy physics to low-energy systems, in ultracold atomic gases. Paradigmatic examples are provided by the hard-sphere Bose gas~\cite{bosegas1,bosegas2}, and quasi-one-dimensional Bose gas~\cite{qbosegas1,qbosegas2,qbosegas3}. Analogous $k$-body interacting terms, where $k$ is arbitrary, further appear in fractional quantum Hall systems, where the projection to a single Landau level and the incorporation of virtual transitions to higher levels result in multibody effects~\cite{llevel1,llevel2,llevel3}. In the context of ultracold bosonic atoms confined in optical lattices, such multibody contributions can be spectroscopically resolved~\cite{spect} and are responsible for distinctive many-body phenomena, such as the collapse and revival dynamics observed in quantum phase evolution experiments~\cite{col_revival1,col_revival2,col_revival3,col_revival4}. Moreover, there exist certain scenarios, where all multibody terms up to the $k$-body prevail. However, the strength of such interactions can be controlled to realize situations where  the higher-body force dominates over the lesser-body forces~\cite{3body2,manybody1,3body1,3body3,3body4,3body6}. \\

\noindent 
\emph{\uline{Definition of symmetric states.}}- We refer to a state as symmetric if all $l$-party reduced states are equal for every fixed $l$, for $l=1$ to $N$. If a state does not satisfy this property, we refer to it as asymmetric. \\

\noindent 
\emph{\uline{Definition of genuine multipartite entanglement.}}- An $N$-partite pure quantum state is genuine multipartite entangled iff all the bipartite partitions of the state generates reduced density matrices that are mixed in nature~\cite{entrev}. I.e., there is no bipartite partition in which the reduced states are pure. The GHZ state is an example of a genuine multipartite entangled state.

\vspace{0.2cm}

\section{Scaling of QFI in estimating $k$-body interaction strengths (case I)}
We maximize the QFI with respect to all input probes, and find the maximum QFI in the limit of large number of system particles, while estimating coupling strength $J$ for the hamiltonian, $Jh_k^{(N)}$,  where $k$ is arbitrary but fixed. For completeness, we also provide a short analysis of the scaling of QFI for such Hamiltonians, while restricting to symmetric product probes.
\begin{theorem}
\label{th1}
    The maximum quantum Fisher information, optimized over all input probes, which is attainable with arbitrary-body encoding Hamiltonians, where  each single-body term is a Pauli-$\sigma_z$ operator, scales as $N^{2k}$, in the limit $N>>k$. Here $N$ represents the number of parties in the system and $k$ denotes the number of parties interacting with each other.
\end{theorem}

\noindent \textit{Proof.} 
Corresponding to the encoding that we consider in the case I, the QFI is given by $4\Delta^2 h_k^{(N)}$, where $\Delta$ denotes variance.
Given a hermitian operator $A$, its variance,
$\Delta^2 A$, attains the maximum value 
$\Delta^2 A_{max}=(a_M-a_m)^2/4$, where $a_M$ and $a_m$ are the maximum and minimum eigenvalues of $A$ respectively. Further, the state which maximizes the variance is $\ket{\widetilde{\mathcal{A}}}=\left(\ket{a_M}+\ket{a_m}\right)/\sqrt{2}$, where $\ket{a_M}$ and $\ket{a_m}$ are  eigenvectors of $A$ corresponding to the eigenvalues $a_M$ and $a_m$ respectively~\cite{variance}. 
We will use this 
fact
to 
maximize the QFI, given by $4\Delta^2 h_k^{(N)}$.
We prove the theorem separately for the cases when $k$ is even and odd, in the succeeding parts.
\vspace{0.2cm}

\noindent 
\emph{\uline{When encoding Hamiltonian has even-body interactions.}}-
We begin by considering the case where $k$ is even. The maximum variance, and hence the maximum QFI of $h_k^{(N)}$ is given in terms of its maximum and minimum eigenvalues. The maximum eigenvalue of $h_k^{(N)}$ is $\binom{N}{k}$. The corresponding eigenvector is given by $\ket{0}^{\otimes N}$. This is because there are a total of $\binom{N}{k}$ terms in the $k$-body Hamiltonian, and the maximum eigenvalue is obtained when $\ket{0}$ acts locally on each party. 
Next we evaluate the minimum eigenvalue of $h_k^{(N)}$. 
We refer to an arbitrary eigenstate of $h_k^{(N)}$ with $m$ up spins and $N-m$ down spins  as  $\ket{E_m}=\mathcal{P}[\ket{0}^{\otimes m}\otimes \ket{1}^{\otimes N-m}]$, with \(\mathcal{P}\) being an arbitrary permutation of the kets in its argument. Our aim is to find the optimum value of $m$, 
which minimizes the corresponding eigenvalue, 
 $E_m$. 
The Hamiltonian has $\binom{N}{k}$ terms. Each of these terms acts on the state, $\ket{E_m}$, and gives a real number times the state, $\ket{E_m}$. Summing up all such terms, we obtain the eigenvalue of the Hamiltonian, $h_k^{(N)}$, corresponding to the eigenstate, $\ket{E_m}$. First, we cluster all the terms in the Hamiltonian into $k$ groups depending on whether the single-body operators in each term act on $\ket{0}$ or $\ket{1}$ in the eigenstate, $\ket{E_m}$. There are some terms in $h_k^{(N)}$, for which all the single-body operators act on the ket $\ket{1}$, and no operator acts on the state, $\ket{0}$. The number of such terms in the Hamiltonian is $\binom{m}{0}\binom{N-m}{k}$. This gives an eigenvalue $(-1)^k$=1, since $k$ is even. Similarly there are $\binom{m}{1}\binom{N-m}{k-1}$ terms in the Hamiltonian in which only $1$ term acts on the ket, $\ket{0}$, while the rest act on $\ket{1}$, generating an eigenvalue $(-1)^{k-1}$=-1. The contribution of all the $\binom{N}{k}$ terms to the eigenvalue when they act on the state, $\ket{E_m}$, is obtained by adding all these numbers, $(-1)^i\binom{m}{i}\binom{N-m}{k-i}$, where $i$ runs from $1$ to $k$. So the eigenvalue of $h_k^{(N)}$, when it acts on the state $\ket{E_m}$ is given by
    $\lambda_m=\sum_{i=0}^{k} (-1)^i \binom{m}{i}\binom{N-m}{k-i}$.

Now consider the equation, $\sum_{i=0}^{k/2} \binom{m}{2i}\binom{N-m}{k-2i}=( \widetilde{\lambda}+\lambda_m)/2$, where $\widetilde{\lambda}=\binom{m}{i}\binom{N-m}{k-i}$. The quantity, $\widetilde{\lambda}$ is the coefficient of $x^k$ in the expansion of $(1+x)^m(1+x)^{N-m}$. This coefficient is simply given by $\binom{N}{k}$. This gives $\lambda_m=2\sum_{i=0}^{k/2} \binom{m}{2i}\binom{N-m}{k-2i}-\binom{N}{k}$. Therefore the quantity, $(\binom{N}{k}-\lambda_m)^2$ is given by $4\left[\binom{N}{k}-\sum_{i=0}^{k/2} \binom{m}{2i}\binom{N-m}{k-2i}\right]^2=4[\binom{N}{k}-\widetilde{\lambda}_m]^2$, where $\widetilde{\lambda}_m=\sum_{i=0}^{k/2} \binom{m}{2i}\binom{N-m}{k-2i}$.
Since $m$ is a positive integer, \(\widetilde{\lambda}_m\), as a function of \(m\) has a discrete domain. For large \(N\), we approximate it as a continuous function values of $\widetilde{\lambda}_m$ for all $m$. Assuming the continuity, the minimum eigenvalue of $h_k^{(N)}$ is $\min_m \widetilde{\lambda}_m$. We call the optimal $m$ after the minimization $m_0$. In order to find the optimal feasible integer value of $m$, we have to consider the integer nearest to $m_0$. 
Therefore the maximum QFI in this scenario, when optimized over all input states is given by
\begin{eqnarray}
    F^{opt}_{e}=4\left[\binom{N}{k}-\min_m \sum_{i=0}^{k/2} \binom{m}{2i}\binom{N-m}{k-2i}\right]^2,
\end{eqnarray}
where $m$ is any integer from $0$ to $N$.

The exact QFI for $k=2$ is provided in the Appendix.
For higher values of $k$, we find the scaling of the maximum QFI with $N$ by performing an exact analytical analysis in the limit $N\gg k$. 
For details of the calculations, refer to Appendix~\ref{even}~\cite{descartes1,descartes2,descartes3,descartes4,descartes5}.
Utilizing an entirely analytical approach, we find that, in the limit $N\gg k$, the only positive real value of $m$, which corresponds to the minimum of $\widetilde{\lambda}_m$ is at $m=N/2$. Therefore the maximum QFI is given by 
\begin{eqnarray}
    F_{opt}^e = \left[\binom{N}{k}-(-1)^{k/2}\binom{N/2}{k/2}\right]^2.
\end{eqnarray}
The quantity, $F_{opt}^e$, can be further simplified using Stirling approximation in the limit $N \gg k$. 
In such a limit, it follows that
the maximum QFI for even-body interactions 
scales as  $F_{opt}^e \sim N^{2k}/(k!)^2$.
It is to note that, since the equation, $\partial \widetilde{\lambda}_m/\partial m=0$, has only one positive real solution at $m=N/2$ in the limit $N\gg k$, the minimum value of the function, $\widetilde{\lambda}_m$, at $m=N/2$ is indeed a global minimum in the limit of large values of $N$.
We have also backed our analytics by an alternative semi-analytic approach for optimization, 
 the details of which are provided in Appendix~\ref{even}~\cite{lsqft1,lsqft2}.  \vspace{0.2cm}

\noindent 
\emph{\uline{When encoding Hamiltonian has odd-body interactions.}}-
A possible choice of the maximum eigenvalue for odd values of $k$ corresponds to the eigenvector with all the local spins pointing upwards, i.e. $\ket{0}^{\otimes N}$, with eigenvalue $\binom{N}{k}$.
Further, since there are an odd number of terms in the Hamiltonian, the minimum eigenvalue would be $-\binom{N}{k}$ with  $\ket{1}^{\otimes N}$ as the corresponding eigenvector. Therefore the maximum QFI for $k$-body interactions involving $N$ parties when $k$ is odd is given by
    $F_{opt}^o=4\left\{\binom{N}{k}\right\}^2$.
In the limit $N\gg k$, 
the quantity, $F_{opt}^o$, scales as $\sim 4N^{2k}/(k!)^2$.\\

\noindent 
This completes the proof of Theorem~\ref{th1}. \hfill $\blacksquare$

\vspace{0.2cm}

 \noindent 
\emph{\uline{Scaling of QFI for optimum symmetric product probes for arbitrary-body encoding Hamiltonians.}}-
We scrutinize the minimal error obtained in the estimation of coupling strength of arbitrary-body interactions, by maximizing the relevant QFI over product input probes, which are symmetric in nature.
We find that in this case, the maximum QFI scales as $\sim N^{2k-1}$, in the limit $N\gg k$. For the details of the calculation, refer to Appendix~\ref{sym_prod}. \vspace{0.2cm}
\section{Scaling of QFI in estimating up to $k$-body interaction strengths (case II)}
\begin{figure}
\centering
\includegraphics[width=8cm]{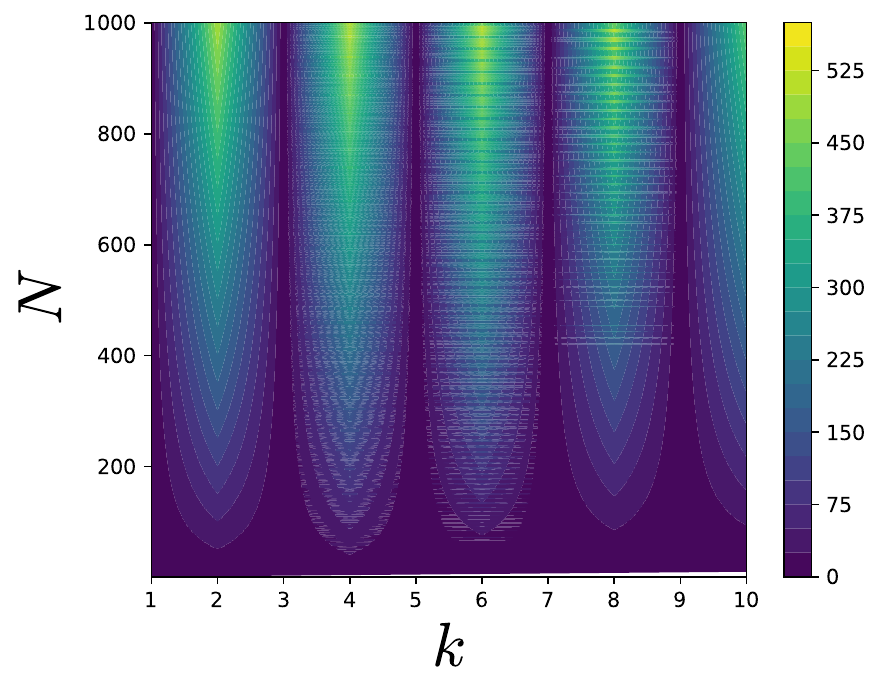}%
\caption{The optimal value of $m$ (refer to Eq.~\eqref{uptok}), which corresponds to the maximum QFI, when the encoding Hamiltonian has up to $k$-body interactions, i.e. considering case II. $N$ denotes the total number of parties. The figure depicts that for odd values of $k$, the maximum QFI is attained at $m=0$, while for even $k$, it is not so.}
\label{upto_k_body}
\end{figure}

The maximum QFI in this scenario is given by $4\Delta^2 K_x$, where $K_x=\sum_{i=1}^k ix^{i-1}h_i^{(N)}$. Our aim is to find out if the dichotomy observed previously in the nature of the optimal states also prevails here. The state that maximizes the QFI is given by $\ket{\Psi}=(\ket{\lambda_{max}^K}+\ket{\lambda_{min}^K})/\sqrt{2}$, where $\ket{\lambda_{max(min)}^K}$ is the eigenvector corresponding to the maximum (minimum) eigenvalue of $K_x$. Following an argument analogous to the case I, the maximum eigenvector is given by $\ket{0}^{\otimes N}$. Now, the eigenvalue of $K_x$ corresponding to an arbitrary eigenstate, $\mathcal{P}[\ket{0}^{\otimes m} \ket{1}^{\otimes N-m}]$, is given by
\begin{eqnarray}
\label{uptok}
    \lambda_m^K=\sum_{k_1=1}^{k}\left(\sum_{i=0}^{k_1}\binom{m}{k_1-i}\binom{N-m}{i}(-1)^i\right)k_1 x^{k_1-1}, \nonumber
\end{eqnarray}
with \(\mathcal{P}\) being an arbitrary permutation of the kets in its argument.
In order to find the optimal state that maximizes the QFI, we minimize $\lambda_m^K$ with respect to $m$. In Fig.~\ref{upto_k_body}, we depict the behavior of the optimal value of $m$ as functions of $k$ and $N$. We find that for odd values of $k$, the minimum is at $m=0$, while for $k$ being even, $m=0$ does not correspond to the minimum. This has been further verified numerically by checking that $\lambda_{min}<\lambda_m^K(m=0)$ for $k$ being even. This proves that for encoding Hamiltonians with up to even-body interactions, the optimal probe is asymmetric and not genuine multiparty entangled, while that for up to odd-body interactions is the highly symmetric genuine multipartite entangled GHZ state.

\vspace{0.2cm}

\noindent 
\section{Entanglement vs. asymmetry in optimal probes for even- and odd-body interactions}
Here we find some intriguing features of the optimal probes, separately for even- and odd-body interacting encoding Hamiltonians, in the estimation of coupling constant. Let us first consider the even-body case. 



\vspace{0.2cm}

\noindent 
\emph{\uline{Even-body interactions.}}- When $k$ is even, one of the eigenvectors of the Hamiltonian, $h_k^{(N)}$, 
is given by $\ket{E_{m}}=\ket{0}^{\otimes m}\otimes \ket{1}^{\otimes N-m}$. 
The optimal value of $m$, i.e. $m_0$ which minimizes the eigenvalue, $E_m$, corresponds to the eigenstate $\ket{E_{m_0}}$.
It is argued in Appendix~\ref{amaltas} that $m=0$ would imply a fully degenerate Hamiltonian, which is clearly a contradiction, since the Hamiltonian consists of tensor products of Pauli-$z$.
Further, the eigenvector corresponding to the maximum eigenvalue for $k$-body interactions, where $k$ is even, is given by $\ket{E_{max}}=\ket{0}^{\otimes N}$.
Therefore, since $m\ne 0$,  the optimum eigenstate is never genuine multiparty entangled.
The case of $k=2$ is discussed in detail in Appendix~\ref{amaltas}.
Specifically, the optimum states for $k=2$, and for even and odd values of $N$ are given by
    $ \ket{\chi_e}=\frac{1}{\sqrt{2}}\left(\ket{0}^{\otimes N/2}+\ket{1}^{\otimes N/2}\right)\otimes \ket{1}^{\otimes N/2} $,  and
     $ \ket{\chi_o}=\frac{1}{\sqrt{2}}\left(\ket{0}^{\otimes (N-1)/2}+\ket{1}^{\otimes (N-1)/2}\right)\otimes  \ket{1}^{\otimes (N+1)/2}$ 
respectively.

\vspace{0.2cm}

\noindent 
\emph{\uline{When is the optimal probe an asymmetric product?.}}- There exist possible choices of optimal probes for $k=2$ and $N=2,3$ which are asymmetric product. Motivated by this intriguing feature for $k=2$, we delve deeper to find whether this kind of asymmetry exists in $k$-body interactions with higher values of $k$ where $k$ is even. Let us consider, an $N$-party state where there are $m$ up spins $(\ket{0})$ and $N-m$ down spins  $(\ket{1})$, i.e.
    $\Psi=\ket{0}^{\otimes m}\otimes\ket{1}^{\otimes {N-m}}$
The minimum eigenvalue corresponding to this state is $\lambda(m)=\min_m 2n_0(m)-\binom{N}{k}$. Since $m$ is a positive integer, \(\lambda\), as a function of \(m\) has a discrete domain. For large \(N\), we approximate it as a continuous function values of $\lambda(m)$ for all $m$. Assuming the continuity, the minimum eigenvalue of the \(k\)-body generator is $\lambda(m)=\min_m 2n_0(m)-\binom{N}{k}$. We call the optimal $m$ after the minimisation to be $m_0$. In order to find the optimal feasible integer value of $m$, we have to consider the integer nearest to $m_0$. Further,  the corresponding eigenstate is asymmetric product if the optimal $m$, i.e. $m_0$, is equal to $1$. As, we considered the closest integer of $m_0$, the relevant maximum value of $N$ for which $0.5<m_0<1.5$, will give the maximum number of parties up to which  asymmetric product state is a possible choice of optimal input probe. Performing numerical analysis we obtained a range of $N$, in which asymmetric product state will always be optimal for $k$-body generators, where $k$ is even. This range is given by
  $ k \leq N \le 2k-1 $,
for even values of $k$ starting from $2$. Let the corresponding maximum $N$ be denoted by $N_{max}$. For two-body interactions, i.e. $k=2$, this implies $N_{max} = 3$. This exactly matches our observation discussed previously. 

\vspace{0.2cm}

\noindent 
\emph{\uline{Odd-body interactions.}}- We found that one possible choice of the maximum eigenvalue for odd values of $k$ corresponds to the eigenvector with all the local spins pointing upwards, i.e. $\ket{0}^{\otimes N}$. Similarly the eigenvector corresponding to the minimum eigenvalue is $\ket{1}^{\otimes N}$. Therefore one can argue that the optimal eigenstate corresponding to $k$-body Hamiltonians, when $k$ is odd, always possess some non-zero amount of genuine multipartite entanglement. In particular, the optimal state is of the form, $(\ket{0}^{\otimes N}+\ket{1}^{\otimes N})/\sqrt{2}$.

\vspace{0.2cm}
%
\vspace{0.2cm}

\noindent 
\section{Higher dimensional probes}
Here we present the optimal states for two- and three-body encoding Hamiltonians whose each local component has maximum (minimum) eigenvalue $\delta_{M(m)}$.
The dichotomy that we found in the two-dimensional case is similar to the scenario, $\mathcal{A}_5$, given in Table~\ref{table1}. This shows that this dichotomy is present even in dimensions greater than two.
A detailed discussion on higher dimensional probes is provided in the SM. 
\begin{table}[!htb]
\centering
\begin{tabular}{|c|c|c|c|}
\hline
    & Scenario & 2-body  &  3-body    \\  
  \hline
  \hline
  $\mathcal{A}_3$ & $\delta_M >0$, $\delta_m <0$, $|\delta_M|>|\delta_m|$ & $\ket{\delta_M}\ket{\phi_2}$  & $\ket{\delta_M}^{\otimes 2}\ket{\phi_2}$  \\ 
   \hline
   $\mathcal{A}_4$ & $\delta_M >0$, $\delta_m <0$, $|\delta_M|<|\delta_m|$ & $\ket{\delta_m}\ket{\phi_2}$ &  $\ket{\delta_m}^{\otimes 2}\ket{\phi_2}$ \\ 
   \hline
  $\mathcal{A}_5$ & $\delta_M >0$, $\delta_m <0$, $|\delta_M|=|\delta_m|$ & $\ket{\delta_{M(m)}}\ket{\phi_2}$  &  $\ket{\phi_4}$ \\ 
   \hline
   \end{tabular}
\caption{Optimum states that maximize the QFI for two- and three-body encoding Hamiltonians having three and four parties respectively. The state $\ket{\phi_s}=(\ket{\delta_m}^{\otimes s}+\ket{\delta_M}^{\otimes s})/\sqrt{2}$, for any value of $s$.}
\label{table1}
\end{table}
\vspace{0.2cm}

\noindent 
\section{Conclusion}
In this letter, we dealt with the problem of quantum parameter estimation when the encoding Hamiltonian consists of $k$-body interactions, where $k$ is arbitrary. Specifically, we considered two types of probes - optimal symmetric and optimal ones. In both the cases, we found that in the limit of large system particles compared to $k$, the scaling of quantum Fisher information increases monotonically with the number of particles. No apparent dichotomy was found for even- and odd-body interactions in the scaling of quantum Fisher information for large system particles. The cases corresponding to $k=2$ and $3$ have been provided special forms. Comparing the cases with respect to the scaling of QFI in the  limit of large number of probes, we found that allowing arbitrary input states leads to better scaling than the symmetric product one.

For even-body interactions, the optimum input states proved to be ones that does not possess any genuine multipartite entanglement, but must be asymmetric. This is in drastic contrast with the one-body generator case, where the best optimal  state must be genuine multipartite entangled. This leads us to conclude that as the interaction in metrological encoding is increased to $k$-body, with $k$ being even, asymmetry and not genuine multiparty entanglement is a resource for better precision in quantum parameter estimation. Strikingly, as we shifted to odd-body interactions, the GHZ state, which possesses a non-zero GME, proved to be the optimum one for any number of parties. We therefore have a dichotomy with respect to the presence and absence of genuine multipartite entanglement for even and odd $k$, respectively, when $k$-body interactions are utilised in the generator of the encoding unitary in quantum parameter estimation, with the probe being of an arbitrary number of parties. We thereby find a complementarity in resource requirement in optimal probes for estimating strength of odd- and even-body interactions. Specifically, for odd-body interactions, genuine multiparty entanglement is essential in the optimal probes, but asymmetry in such states is not important. However for even-body interactions, asymmetric probes are useful in attaining optimality, though genuine multipartite entanglement is not necessary in the optimal probes in this scenario. We further extend our analysis to Hamiltonians comprising local fields and up to $k$-body interaction terms, where the interaction strength diminishes progressively with an increase in interaction strength. Notably, we observe a similar dichotomous behavior in the structure of the optimal probe states, depending on whether the Hamiltonian includes interactions up to even- or odd-body terms.

Further, we provide a bound on the number of parties up to which one can always obtain an asymmetric product state as an optimum probe for even values of $k$. Finally, we have also considered two- and three-body interaction with arbitrary local dimension, for certain number of subsystems, utilizing the feature that the maximum quantum Fisher information can be fully characterized by the maximum and minimum eigenvalues of the local component of the Hamiltonian. We have obtained the optimum input in this scenario, and found that the optimum input probe can have zero or non-zero genuine multipartite entanglement depending upon the relative signs of the the maximum and minimum eigenvalues of the local component of the  Hamiltonian. This leads us to establish conditions on maximum and minimum eigenvalues of the local components of the Hamiltonian that guarantees benefits with asymmetric probe states.


\appendix
\section{Preliminaries}
\label{prelim}
In this section, we discuss the quantum parameter estimation protocol in general,
and then provide a description of the encoding Hamiltonians relevant for our analyses.
\subsection{Quantum Fisher information and parameter estimation protocol}
\label{sec1}
A parameter $\xi$ is encoded onto an input probe by a physical process, and is followed by performance of a suitable measurement on the encoded probe. From the knowledge of the measurement outcomes, one can employ a particular estimator to estimate the unknown parameter $\xi$. Our goal is  to calculate the minimum error in the estimation of the encoded  parameter, $\xi$. 
Let the measurement operator be \(M_x\), and let \(x\) be an outcome corresponding to the same. 
Here, 
$M_x>0$ and $M_x^{\dagger}=M_x$ \(\forall x\), and  $\sum_{x}M_{x}=\mathcal{I}$, the identity operator on the Hilbert space corresponding to the physical system at hand.
Given an outcome, $x$, we try to estimate the value of $\xi$, 
depending on
an estimator function, $g(x)$.
For a specific $\xi$, an estimator function whose average over the measurement outcomes gives the true value of the estimated parameter, is called an unbiased estimator~\cite{Braunstein1,review1}. If the measurement outcomes, $x$, belong to a probability distribution, $f(x|\xi)$, then the condition of unbiasedness is given by 
\begin{equation}
    \langle g(x) \rangle_\xi := \int dx f(x|\xi) g(x) = \xi. 
    \label{brisTi}
\end{equation}
Since the standard deviation of the estimator function determines 
an error in estimation, we aim at minimising the error by choosing the best estimator function, defined as one which has the minimum standard deviation. 

Quantum metrology provides a  lower bound to the standard deviation of an unbiased estimator, which is referred to as the Cram\'er-Rao bound~\cite{holevo}, 
and is given by
\begin{equation}
\label{ClCramerRao}
    \Delta \xi \geq \frac{1}{\sqrt{\nu F(\xi)}}.
\end{equation}
Here we have used the customary notation $\Delta \xi$ for $\sqrt{\Delta^2 g(x)}$, the standard deviation of the distribution of the estimator $g(x)$.
 The quantity, $\nu$, denotes the number of times the entire process of estimation is repeated, and $F(\xi)$ is the total Fisher information. 
In ineq.~\eqref{ClCramerRao}, the
definition of Fisher information, $F(\xi)$, is given in terms of the probability distribution, $f(x|\xi)$, by
\begin{eqnarray}
F(\xi) &=& \int dx f(x|\xi) \big[\frac{\partial}{\partial \xi}\log f(x|\xi)\big]^2.
\end{eqnarray}
Clearly, $F(\xi)$ is independent of $g(x)$,
implying that minimum error in estimation of $\xi$ can be discerned, without the knowledge of the best  estimator function, as long as the probability of the measurement outcome, $f(x|\xi)$, is known for every  $x$.  The estimator is still to be unbiased, however. 
In general, Fisher information may depend on the parameter, $\xi$,
and hence the notation, $F(\xi)$. 
The Fisher information also depends, in general, on the choice of the input probe and the measurement performed.
Therefore in ineq.~\eqref{ClCramerRao}, for a given input state, the error can be further minimised by maximising the Fisher information, $F(\xi)$, over all possible measurements. The quantity, hence obtained, is termed as the quantum Fisher information (QFI), denoted by $F_{Q}(\xi)$. The optimal choice of measurements which maximises $F(\xi)$ can also be identified~\cite{Braunstein1}. As a consequence of this optimisation, one arrives at the quantum Cram\'er-Rao bound, which is given by the rightmost term in the following inequality: 
\begin{equation}
\label{QCRbound}
\Delta \xi \geq  \frac{1}{\sqrt{ \nu F(\xi)}} \geq  \frac{1}{\sqrt{ \nu F_Q(\xi)}}. 
\end{equation}
This bound can be achieved by performing a projective measurement in 
the eigenbasis of an operator known as the symmetric logarithmic derivative (SLD)~\cite{Braunstein1}.
The symmetric logarithmic derivative, $L_s$, corresponding to an encoded state, $\rho(\xi)$, is given in terms of the following equation:
\begin{equation}
    \frac{\partial \rho(\xi)}{\partial \xi}=\frac{1}{2} \left[L_s \rho(\xi)+\rho(\xi) L_s \right].
\end{equation}
The QFI can be explicitly expressed in terms of the encoded state, $\rho(\xi)$, using the 
relation 
$F_Q(\xi)=\text{Tr}\left[\rho(\xi) {L_s}^2 \right]$.

Further simplifications in the expression of QFI is obtained if we restrict to pure input states and consider only unitary encodings. Since the operation is unitary, the encoded state is also pure, which we denoted by $\ket{\psi(\xi)}$. 
In such a scenario, the QFI  reduces to
$F_Q(\xi)=4 \big[\braket{\dot{\psi}(\xi)|\dot{\psi}(\xi)}-|\braket{\dot{\psi}(\xi)|\psi(\xi)}|^2 \big]$,
where the dots represent derivatives with respect to $\xi$. 
Let the encoding process be governed by a unitary evolution given by \(\exp(-i\kappa \tilde{h}t/\hbar)\), where \(\kappa\) and \(t\) have the units of energy and time respectively, whereby the parameters in the Hamiltonian \(\tilde{h}\) are dimensionless. 
The dimensionless quantity \(\kappa t/\hbar\) is set to unity, which means that we perform the measurement at time \(\hbar/\kappa\). 
So the parameter, $\xi$, that is encoded in the probe state by the  operator, $\tilde{h}$, is dimensionless. 
The action of the encoding unitary, $U_{\xi}$, is therefore given by
\begin{equation}
\label{unitary}
U_{\xi}\ket{\psi_0}=e^{-i\tilde{h}(\xi)}\ket{\psi_0}=\ket{\psi(\xi)},
\end{equation}
 where $\ket{\psi_0}$ is the input state comprising of $N$ probes and $\ket{\psi(\xi)}$ denotes the corresponding encoded state.
 The expression of QFI,
 corresponding to the encoding given in Eq.~\eqref{unitary},
 simplifies to $F_Q(\xi)=4\Delta^2h$, provided the Hamiltonian, \(\tilde{h}\), is given by 
 \(\tilde{h}= \xi h\), where the operator \(h\) is independent of \(\xi\). 
The quantity, $\Delta^2h$, denotes the variance of $h$ 
in
the state $\ket{\psi(\xi)}$, or equivalently 
in
$\ket{\psi_0}$.
In this framework, since $[h,U_{\xi}]=0$, the QFI becomes independent of the parameter that we want to estimate. Hence we drop the notation $\xi$ in the designation of QFI and call it $F_Q$. 
The choice of input probe, however, is kept arbitrary in this entire analysis, and $F_Q$ depends on the initial state chosen, for a given generator.
In principle, one can also minimise $\Delta \xi$ with respect to the input state to obtain the minimum error in the estimation of $\xi$. In such cases, $\Delta h$ has to be maximised over the choice of input probes.
So for pure states encoded unitarily following Eq.~\eqref{unitary}, the quantum Cram\'er-Rao bound becomes
\begin{equation}
\label{bound}
    \Delta \xi \ge \frac{1}{2\sqrt{\nu}\Delta h}.
\end{equation}
Here the notation, $\Delta h=\sqrt{\Delta^2 h}$, has been used.
In this paper, we concentrate only on unitary encodings of pure input states, but solely consider the effect of interaction terms in the generator.
 
\subsection{Optimal probe for encoding Hamiltonian with single-body terms}
Before analyzing interactions between different parties, let us briefly recall the behaviour of QFI obtained for Hamiltonians having only local terms, i.e., the unitary operator acting locally on each subsystem. Such a Hamiltonian can be written as $h_1^{(N)}=J\sum_{j=1}^N H_j$, where each local term, $H_j$, acts on the $j^{\text{th}}$ subsystem. $N$ denotes the total number of parties involved. Henceforth, except in Sec.~\ref{higher}, we consider each party to be a qubit, and so each local Hamiltonian, \(H_j\), can, without loss of generality, be considered as a Pauli-\(\sigma_z\) operator. If our aim is to calculate the minimum error in the estimation of field strength, $J$,
we obtain the QFI, $F_Q=4\Delta^2h$, and one can further maximize the QFI with respect to different choices of input states. In this context, two types of pure input probes can be considered. In the first type, we can constrain the input to be a product over all subsystems, a restriction that may well be reasonable in resource-limited situations - no nonlocal operation is needed to create them. 
In the second case, there is no restriction on the probe used, and arbitrary inputs are allowed, and is reasonable in situations where resource (specifically, nonlocal operations) is not in short supply. 
In the seminal paper by Maccone \emph{et  al.}~\cite{Maccone1}, it was shown that if one considers Hamiltonians of the type $h_1^{(N)}$, the  optimum input state is always a genuine multipartite entangled state, the Greenberger-Horne-Zeilinger (GHZ) state~\cite{ghz1,ghz2}, and is of the form, $\ket{\widetilde{\psi}}_e=( \ket{0}^{\otimes N} + \ket{1}^{\otimes N})/\sqrt{2}$, where $\ket{0}$ and $\ket{1}$ are the eigenvectors corresponding to maximum and minimum eigenvalues of $\sigma_z$. The optimum product input state, on the other hand, is given by $\ket{\widetilde{\psi}}_p=\left[ \left( \ket{0} + \ket{1} \right)/\sqrt{2} \right]^{\otimes N}$. The important observation here is that the optimum product input, $\ket{\widetilde{\psi}}_p$, is symmetric with respect to the $N$ parties, and the optimum input probe, i.e. $\ket{\widetilde{\psi}}_e$, has non-zero genuine multipartite entanglement. 


The maximum attainable QFI for local Hamiltonians with  optimal product probes is  $F_1^p=N$, and the maximum QFI achievable for local Hamiltonians is given by $F_1^e=N^2$. The maximal QFI in this situation corresponds to the genuine multiparty entangled states,  $\ket{\widetilde{\psi}}_e$.  In Ref.~\cite{Maccone1}, the uncorrelated input was referred to as ``classical", and the genuinely multiparty  entangled GHZ input states as ``quantum" states. 
The minimum error in the estimation of $J$ provides a scaling $\sim 1/\sqrt{N}$, often referred to as the standard limit or shot noise limit (SNL), if the input state is classical. Whereas, if the input state is quantum, a scaling $\sim 1/N$ is produced, which is referred to as the Heisenberg limit (HL).

\section{Scaling of minimum error in estimating coupling strength for arbitrary-body interactions}
\label{jyotsna}
We find the scaling of minimum error in the estimation of coupling strength $J$ for two types of probes, viz. optimal  probes and optimal symmetric ones, while the encoding Hamiltonian comprises $k$-body interactions, where $k$ is arbitrary but fixed. In the first two subsections, we consider optimal probes, where we derive the exact analytical scaling of the QFI with the number of particles in the limit of large number of particles and arbitrary $k$, while in the third subsection, we deal with optimal symmetric product probes. We are particularly interested in the domain $N\gg k$, where $N$ denotes the number of parties. Along the way, situations corresponding to $k=2$ and $3$ are given special 
attention.

\begin{figure}
\centering
\includegraphics[width=8cm]{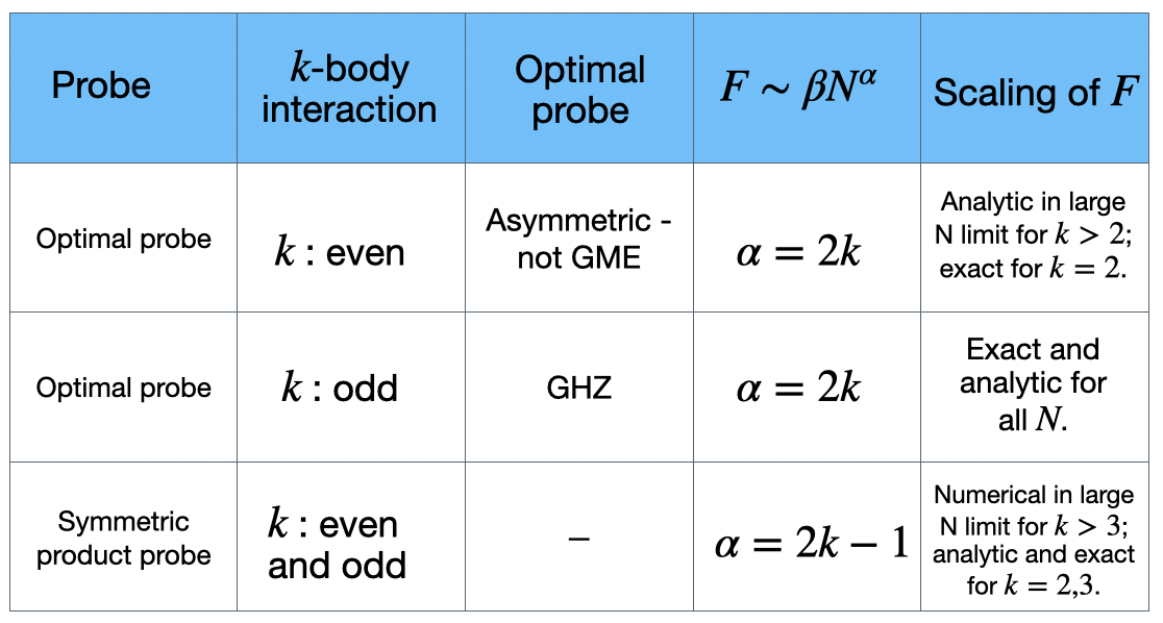}%
\caption{We present here a summary of the different cases considered in the paper and a gist of the results obtained. The quantities, $F$ and $N$, denote the QFI and the number of parties respectively, while $\alpha$ and $\beta$ are the scaling factors.}
\label{table}
\end{figure}
%
%
\subsection{When encoding Hamiltonian has even-body interactions}
\label{even}

We begin by considering the case when $k$ is even. The maximum variance of $h_k^{(N)}$ is given in terms of its maximum and minimum eigenvalues. The maximum eigenvalue of $h_k^{(N)}$ is $\binom{N}{k}$. The corresponding eigenvector is given by $\ket{0}^{\otimes N}$. This is because there are a total of $\binom{N}{k}$ terms in the $k$-body Hamiltonian, and the maximum eigenvalue is obtained when $\ket{0}$ acts locally on each party. 

Next we evaluate the minimum eigenvalue of $h_k^{(N)}$. 
Let us suppose that there are $m_0$ up spins $(\ket{0})$ and $N-m_0$ down spins $(\ket{1})$ in the eigenstate corresponding to the minimum eigenvalue. We refer to an arbitrary eigenstate of $h_k^{(N)}$ with $m$ up spins and $N-m$ down spins  as  $\ket{E_m}=\mathcal{P}[\ket{0}^{\otimes m}\otimes \ket{1}^{\otimes N-m}]$, with \(\mathcal{P}\) being an arbitrary permutation of the kets in its argument. Our aim is to find the optimum value of $m$, i.e. $m_0$, which minimizes the corresponding eigenvalue, 
 $E_m$. 
Notice that $\ket{E_m}$ is an  eigenstate of each term in the expansion of $h_k^{(N)}$ with eigenvalues $+1$  or $-1$. 
Now the eigenvalue of a single term in the expansion of $h_k^{(N)}$ is $+1$ when there are an even number of up spins or down spins in the corresponding eigenstate. 
This can occur in $n_0(m)$
ways, where
\begin{eqnarray} n_0(m)=\sum_{i=0}^{k/2}\binom{m}{2i}\binom{N-m}{k-2i}.
 \label{chhaya_neer}
\end{eqnarray}
The number of possibilities in which $-1$ can appear in the eigenvalue of a single term in the expansion of $h_k^{(N)}$ is $\binom{N}{k}-n_0(m)$. So the minimum eigenvalue in this case is $\min_m 2 n_0(m)-\binom{N}{k}$. Therefore the maximal QFI corresponding to the Hamiltonian, $h_k^{(N)}$, where $k$ is even, is given by 
\begin{equation}
    F_{opt}^e=4 \left[\binom{N}{k}-\min_m n_0(m) \right]^2.
\end{equation}

\subsubsection{Exact analysis for $k=2$}
Let us begin by considering the case of $k=2$. 
The minimum eigenvalue of $h_2^{(N)}$ in that case is 
\begin{eqnarray}
    E_m &=& \min_m \left[\binom{N}{2}-\left\{\binom{m}{2} + \binom{N-m}{2}\right\}\right] \nonumber \\
        &=& \min_m \frac{1}{2}\left((N-2m)^2-N\right).
\end{eqnarray}
Since $m$ is a positive integer, $E_m$ has a discrete domain. But for large $N$, 
we 
approximate it as a  
function on a continuous domain.
Assuming this continuity, $E_m$ can be differentiated with respect to $m$ to find the minimum. The value of $m$ corresponding to the minimum is $\widetilde{m}=N/2$. For odd values of $N$, one has to consider the integer nearest to $N/2$. Thus for odd values of $N$, $\widetilde{m}$ is given by  $(N-1)/2$ or $(N+1)/2$.
The minimum eigenvalue in this case can in general be expressed as $E_m=((1-(-1)^N)/2-N)/2$, for $k=2$ and arbitrary $N$. The maximum  QFI for Hamiltonians having two-body interaction terms involving $N$ parties is therefore given by
\begin{eqnarray}
\label{keq2}
    F_2^{max} &=& \left[\binom{N}{2}-\frac{(1-(-1)^N)/2-N}{2}\right]^2 \nonumber \\
    &=& \frac{1}{16} \left(2N^2-1+(-1)^N\right)^2.
\end{eqnarray}
The $max$ in the superscript implies that this is the maximum attainable QFI  with two-body interactions involving $N$  parties, since we have optimized over all possible input probes.
From Eq.~\eqref{keq2}, we find that the maximal QFI for two-body interactions scales with $N$ as $\sim N^4/4$.

\subsubsection{Analytical method of optimizing the QFI for arbitrary even-body interactions in the limit of large $N$}
We rewrite the quantity $\widetilde{\lambda}_m=\sum_{i=0}^{k/2}n_i$, where $n_i=\binom{m}{2i}\binom{N-m}{k-2i}$.
We can further write
\begin{eqnarray}
\label{ni}
    n_i &=& \frac{\prod_{i'=0}^{2i-1}(m-i')}{(2i)!} \frac{\prod_{i'=0}^{k-2i-1}(N-m-i')}{(k-2i)!} \nonumber \\
        &=& \frac{N^k}{(2i)!(k-2i)!} \prod_{i'=0}^{2i-1} (\frac{m}{N}-\frac{i'}{N}) \prod_{i'=0}^{k-2i-1} (1-\frac{m}{N}-\frac{i'}{N}) \nonumber \\
        &\approx& \frac{N^k}{(2i)!(k-2i)!} \left(\frac{m}{N}\right)^{2i} \left(1-\frac{m}{N}\right)^{k-2i},
\end{eqnarray}
where in the last we have used the limit $N\gg k$.
Taking derivative of both sides of Eq.~\eqref{ni}, and summing over $i$, we obtain
    $\partial \widetilde{\lambda}_m/\partial m = \sum_{i=0}^{k/2}m^{2i-1}(N-m)^{k-2i-1}(2iN-mk)$.
Equating the above equation to zero, we get the following equation
\begin{eqnarray}
\label{solve0}
    2N\sum_{i=0}^{k/2}ix^i=mk \sum_{i=0}^{k/2} x^i,   
\end{eqnarray}
where  $x=m^2/(N-m)^2$.
The summation on both sides of the above equation can be performed. In a compact form, this is given by
\begin{eqnarray}
\label{solve}
    2N\frac{x\left[1-(\frac{k}{2}+1)x^{k/2}+\frac{k}{2}x^{k/2+1}\right]}{(1-x)^2}=mk\frac{1-x^{k/2+1}}{1-x},
\end{eqnarray}
for $x\ne 1$. For $x=1$, we simply obtain $m=N/2$. Now, let us analyze the case $x \ne 1$.
Eq.~\eqref{solve} can be further rewritten as
\begin{eqnarray}
\label{solve2}
    \frac{ky^{k+4}-2y^{k+3}-(2+k)(y^{k+2}-y^3)+2y^2-ky}{2(y+1)(1-y^2)}=0,
\end{eqnarray}
where $y=\sqrt{x}$. Let us rechristen the numerator on the left-hand side of the above equation as $f_k(y)$. Our aim is to solve the equation $f_k(y)=0$, for $y^2 \ne 1$, i.e. $x\ne 1$. It is to note that, since $x$ is a positive real quantity, we want only positive real solutions of $f_k(y)=0$. Using Descartes' rule of signs~\cite{descartes1,descartes2,descartes3,descartes4,descartes5}, we find that there are three changes of signs in the expression of $f_k(y)$, if we arrange the polynomial in $y$ from higher to lower order in $y$. Therefore, there are three positive real roots of $f_k(y)=0$. Further, we find that $f_k(y=1)=f^{\prime}_k(y=1)=f^{\prime \prime}_k(y=1)=0$, and $f_k^{\prime \prime \prime}(y=1)\ne 0$, which implies that $y=1$ is the only positive real root of the equation, $f_k(y)=0$, with multiplicity three. However, this corresponds to the solution of Eq.~\eqref{solve0} for $x\ne 1$. This proves that there are no real positive solutions of Eq.~\eqref{solve0} for $x\ne 1$. We already found that $x=1$ gives $m=N/2$, which is the optimal value of $m$, corresponding to a maxima or minima. Next, we evaluate $ \partial^2 \widetilde{\lambda}_m/\partial^2 m$ at $m=N/2$ to find whether it corresponds to a maxima or minima. This is given by
\begin{eqnarray}
    \frac{\partial^2 \widetilde{\lambda}_m}{\partial^2 m}\Big|_{m=\frac{N}{2}} 
    = \left(\frac{N}{2}\right)^{k-2} \frac{k}{6} (k^2+3k+2) \ge 0,
\end{eqnarray}
which implies that the function, $\widetilde{\lambda}_m$ has its minimum at $m=N/2$. Therefore the maximum QFI is given by 
\begin{eqnarray}
    F_{opt}^e = \left[\binom{N}{k}-(-1)^{k/2}\binom{N/2}{k/2}\right]^2.
\end{eqnarray}

\subsubsection{Semi-analytic method of optimizing the QFI for arbitrary even-body interactions in the limit of large $N$}
We provide an alternative approach to maximize the relevant QFI.
For higher values of $k$, we find the scaling of maximum QFI with $N$ by first performing an exact analytical analysis in the limit $N\gg k$, and then back it by an extensive numerical method for supporting the results.
In the next part, we provide an analysis of obtaining an exact scaling of the maximal QFI with the number of parties, for arbitrary even $k$ in the limit $N\gg k$. We rewrite the quantity $n_0(m)$ (of Eq.~(\ref{chhaya_neer})) as $(S_1+S_2)/2$, where 
\begin{eqnarray}
    S_1&=&\sum_{i=0}^{k}\binom{m}{i}\binom{N-m}{k-i}, \nonumber \\
    S_2&=&\sum_{i=0}^{k}(-1)^i\binom{m}{i}\binom{N-m}{k-i}.
\end{eqnarray}
Now $S_1$ is the coefficient of $x^k$ in the expansion of $(1+x)^m(1+x)^{N-m}$. This coefficient is simply given by $\binom{N}{k}$. Similarly $S_2$ is the coefficient of $x^k$ of the function, $y=(1-x)^m(1+x)^{N-m}$. This can be rewritten as $\ln{y}=m\ln{(1-x)}+(N-m)\ln{(1+x)}$, 
which in the limit, $|x|<1$, can be expanded as
\begin{eqnarray}
    \ln y &=& (N-2m)\sum_{i=1}^{\infty}\frac{x^{2i-1}}{2i-1} - N\sum_{i=1}^{\infty} \frac{x^{2i}}{2i}. \nonumber
\end{eqnarray}
The quantity, $y$, is further obtained as
\begin{eqnarray}
    y 
    &=& e^{\xi_1} e^{\xi_2} \nonumber \\
    &=& \sum_{i=0}^{\infty}\frac{(\xi_1)^i}{i!} \sum_{j=0}^{\infty}\frac{(\xi_2)^j}{j!}, 
\end{eqnarray}
where $\xi_1=(N-2m)\sum_{i=1}^{\infty}x^{2i-1}/(2i-1)$ and $ \xi_2=  - N\sum_{i=1}^{\infty} x^{2i}/(2i)$.  

The quantity, $\exp(\xi_2)$, essentially simplifies to $(1-x^2)^{N/2}$. So the coefficient of $x^{k-s}$ in the expansion of $\exp(\xi_2)$ is given by $\binom{N/2}{(k-s)/2}(-1)^{(k-s)/2}$, for even values of $s$ and $k$.
Now let us consider the coefficient of $x^s$ in the expansion of $\exp(\xi_1)$, where both $s$ and $k$ are even.
Expanding $\exp(\xi_1)$ in series of $x$, we obtain
\begin{eqnarray}
\label{upasanghar}
    e^{\xi_1} = 1 &+& (N-2m)(x+\frac{x^3}{3}+\frac{x^5}{5}+...) \nonumber \\
    &+& \frac{(N-2m)^2}{2!}(x+\frac{x^3}{3}+\frac{x^5}{5}+...)^2 + ... \;\;\;
\end{eqnarray}
Odd powers of $x$ in $\exp(\xi_1)$ does not contribute to $x^k$ since all powers of $x$ in $\exp(\xi_2)$ are even and $k$ is also even. So there are no powers of $x$ with coefficient $N-2m$ that contribute to $x^k$. Now, the coefficient of $(N-2m)^2$ in Eq.~\eqref{upasanghar}, is given by $\sum_{i=1}^{\infty}x^{2i-1}/(2i-1) \sum_{j=1}^{\infty}x^{2j-1}/(2j-1)/2$. For extracting the coefficient of $x^s$, we set $2(i+j)-2=s$. Thereby the coefficient of $x^s$ in the third term of Eq.~\eqref{upasanghar} is $\sum_{i=1}^{\infty}(N-2m)^2/((2i-1)(s+1-2i))/2$. Similarly the coefficients of $x^s$ in the fourth term of the same equation is $a_1 (N-2m)^3/3!$, where $a_1$ is a constant independent of $x$. So on for the coefficients of higher powers of $(N-2m)$ in Eq.~\eqref{upasanghar}. Therefore the coefficient of $x^k$ in the expression of $y$ is given by
\begin{eqnarray}
\label{sharodiya}
    S_2 &=& (N-2m)^2 \sum_{s=2}^k \binom{\frac{N}{2}}{\frac{k-s}{2}}(-1)^{\frac{k-s}{2}} \nonumber \\
   && \Bigg \{  \sum_{i=1}^{\infty}\frac{1}{2(2i-1)(s+1-2i)} + \sum_{i=1}^{\infty} \frac{(N-2m)^i}{(i+2)!}a_i\Bigg\} \nonumber \\
    &=& (N-2m)^2 f_m,
\end{eqnarray}
where $a_i$, $\forall i$, are constants independent of $m$ and $N$. In order to minimize $n_0(m)$ with respect to $m$, the following equation, given by $ \partial S_2/\partial m = 0$, is to be solved for the value of $m$, and the corresponding condition $ \partial^2 S_2/\partial m^2 >0 $ is to be satisfied at the solution of $m$.  Differentiating both sides of Eq.~\eqref{sharodiya} with respect to $m$, we obtain
\begin{eqnarray}
    \frac{\partial S_2}{\partial m} = -4f_m(N-2m) + (N-2m)^2 \frac{\partial f_m}{\partial m}.
\end{eqnarray}
So setting $\partial S_2/\partial m$ equal to zero gives us a solution $m=N/2=m_0$. Next we prove that $m_0$ indeed corresponds to a minima of the function, $S_2$. The second derivative of $S_2$ with respect to $m$ is given by
\begin{eqnarray}
\label{jaihok}
    \frac{\partial^2 S_2}{\partial m^2}\bigg|_{m_0} &=& 8 f_{m_0} \nonumber \\
    &=& \sum_{s=2}^k \binom{\frac{N}{2}}{\frac{k-s}{2}} \sum_{i=1}^{\infty}\frac{4(-1)^{\frac{k-s}{2}}}{(2i-1)(s+1-2i)}. 
\end{eqnarray}
It has been proved numerically that $\partial^2 S_2/\partial m^2|_{m_0} >0 $, for $2\le k \le N$.
This proves that $m_0=N/2$ corresponds to the minimum of the function, $S_2$. 
Therefore the minimum of $(S_1+S_2)/2$, and hence $n_0(m)$, is given by the functional value of $n_0(m)$ at $m=N/2$ in the limit of large $N$. The maximum QFI in such a situation is then given by
\begin{eqnarray}
    F_{opt}^e &=& 4\left\{\binom{N}{k}-\frac{1}{2}\left[\binom{N}{k}+(-1)^{k/2}\binom{N/2}{k/2}\right]\right\}^2 \nonumber \\
    &=& \left[\binom{N}{k}-(-1)^{k/2}\binom{N/2}{k/2}\right]^2.
\end{eqnarray}
The right hand side of this equation can be further simplified using Stirling approximation in the limit $N \gg k$.
In the limit, $N \gg k$, the quantity, $\binom{N}{k}$ can be approximated by $N^{k/2}/(2^{k/2}(k/2)!)$. For the details of this approximation, refer to the main text.
Therefore the maximum QFI for even-body interactions in the limit of a large number of system particles scales as follows: $F_{opt}^e \sim N^{2k}/(k!)^2$, keeping only highest order term in $N$. For instance, in the case of two-body interactions, the maximal QFI scales as $ N^4/4$. This matches exactly with our previously presented exact analytical result for \(k=2\).
The order of $\beta$ that we obtain here, i.e. $\beta \sim 1/(k!)^2$, also matches with the numerical analyses in the succeeding part of this subsection.

However, it is to note that this analysis proves that $m_0=N/2$ corresponds to a minima of the function, but whether it is a global minima, is yet to be proven.
An extensive numerical analysis shows that $m_0=N/2$ is either the global minimum or very close to the global minimum of the function, $n_0(m)$, for different values of even $k$ in the limit $N\gg k$.
If the maximum QFI varies with $N$ as $F_{k}^{max}\sim \beta N^{\alpha}$ in the large $N$ limit, then the numerically obtained values of $\alpha$ for $k=4,6$ and $8$ are presented in Table~\ref{table:opt} with the appropriate error bars. The errors corresponding to the best-fit curves have been calculated using the least squares method. For details about the error calculation, please refer to Appendix~\ref{andhare_alo}. From Table~\ref{table:opt}, we find that for $k$-body interactions, where $k$ is even, the value of $\alpha \approx 2k$. The values of $\beta$ also varies with $k$. We have also evaluated the values of $\beta$ in this case, which support the analytics in the previous part of this subsection. 

\begin{table}[!htb]
\centering
\begin{tabular}{|c|c|c|}
\hline
   \;\;\; $k$ \;\;\; & \;\;\;\;\;\;\;\;\; $\widetilde{\alpha} \pm \delta$ \;\;\;\;\;\;\;\;\; & \;\;\;\;\; $\mathcal{R}$ \;\;\;\;\; \\  
  \hline
  \hline
  4 & 8.005 $\pm$ 0.00289896 & 0.00513364 \\ 
   \hline
   6 & 12.034 $\pm$ 0.00078858 & 0.00368963 \\ 
   \hline
  8 & 16.039 $\pm$ 0.00146871 & 0.0044122 \\ 
   \hline
   \end{tabular}
\caption{Scaling of the optimal QFI with the number of particles in the limit $N \gg k$. The values of $\alpha$ corresponding to the best-fit functions are given here with the obtained minimum $\chi^2$ error, $\mathcal{R}$, given in Appendix~\ref{andhare_alo}. The quantity, $\widetilde{\alpha} \pm \delta$ represents the $95\%$ confidence levels, 
where $\widetilde{\alpha}$ and $\delta$ denote the maximum likelihood estimate 
and the error bar, respectively, for the fitting parameter, $\alpha$. The convergence of the values of $\widetilde{\alpha}$ 
has been checked at values of $N= 2000$. 
} 
\label{table:opt}
\end{table}


\subsection{Scaling of minimal error for optimum symmetric product probes for arbitrary-body encoding Hamiltonians}
\label{sym_prod}
In this subsection, we find the minimal error obtained in the estimation of coupling strength of arbitrary-body interactions, by maximizing the relevant QFI over product input probes, which are symmetric in nature. 
Let us recall our definition of symmetric states, which says that a state is symmetric if all $l$-party reduced states are equal for every fixed $l$, for $l=1$ to $N$. We consider such an arbitrary symmetric state consisting of $N$ parties, given by $\ket{\psi_0}=\ket{\widetilde{\phi}}^{\otimes N}$, where $\ket{\widetilde{\phi}}=\cos\frac{\theta}{2}\ket{0}+e^{i\phi}\sin\frac{\theta}{2}\ket{1}$. Our aim is to maximize $\Delta^2 h_k^{(N)}$ with respect to $\ket{\psi_0}$. 
In order to perform the maximization, it is convenient to express the variance of $h_k^{(N)}$ in terms of the quantities, $N,k$, and the parameters, $\theta$ and $\phi$. The variance of $h_k^{(N)}$ is  given by $\Delta^2 h_k^{(N)}=\langle(h_k^{(N)})^2\rangle-\langle h_k^{(N)}\rangle^2$, where the angular brackets, in this case, denote expectation value with respect to $\ket{\psi_0}$. 
Let us consider 
each
term in the expansion of $(h_k^{(N)})^2 = J^2\left(\sum_{i_1=1, i_1<j_1...<l_1}^{N} \sigma^{i_1}_z \sigma^{j_1}_z ... \sigma^{l_1}_z\right)$ $\left(\sum_{i_2=1, i_2<j_2...<l_2}^{N} \sigma^{i_2}_z \sigma^{j_2}_z ... \sigma^{l_2}_z\right)$. 
Each term in $(h_k^{(N)})^2$ contains tensor products of $\sigma_z$ and $\sigma_z^2$. 
So there are multiples of $\langle\sigma_z\rangle^{2\alpha} \langle\sigma_z^2\rangle^{k-\alpha}$ in the expansion of $\langle (h_k^{(N)})^2 \rangle$, where $\alpha$ is an integer.
Let us find the coefficient of the $i^{\text{th}}$ term in the expansion of $\langle(h_k^{(N)})^2\rangle$, i.e. the coefficient of  $\langle\sigma_z\rangle^{2\alpha} \langle\sigma_z^2\rangle^{k-\alpha}$ with $\alpha=i$.  
$\langle\sigma_z\rangle^{2}$ can be chosen from \(N\) parties in $\binom{N}{i}\binom{N-i}{i}$ ways, while $\langle\sigma_z^2\rangle$ can be selected in $\binom{N-2i}{k-i}$ ways. Therefore, since $\langle\sigma_z\rangle^2=\cos^{2}\theta$ and $\langle\sigma_z^2\rangle=1$, the \(i\)$^{\text{th}}$ term is given by $\binom{N}{i}\binom{N-i}{i}\binom{N-2i}{k-i}\cos^{2i}\theta$. Next, let us consider the second term of $\Delta^2 h_k^{(N)}$, i.e. $\langle h_k^{(N)}\rangle^2$. The quantity, $\langle h_k^{(N)}\rangle$, 
contains $\binom{N}{k}$ terms, where each term contributes to a $\cos^{k}\theta$, and thereby $\langle h_k^{(N)}\rangle^2=[\binom{N}{k}\cos^{k}\theta]^2$. So the QFI in this situation (viz., in the case of a product state probe) is given by
\begin{eqnarray}
\label{sp}
    F_{SP}&=&4\left[f - \binom{N}{k}^2 \cos^{2k}\theta\right], \;\;\;
    \text{where}  \\
    f&=&\sum_{i=0}^k \binom{N}{i}\binom{N-i}{i}\binom{N-2i}{k-i}\cos^{2i}\theta. \nonumber
\end{eqnarray}
The QFI in this scenario
can be written in a closed form as
\begin{eqnarray}
    F_{SP}=4 \binom{N}{k} \left[\, _2F_1(-k,k-N;1;z)-z^k \binom{N}{k}\right], 
\end{eqnarray}
where $z=\cos^{2}\theta$, and 
$_2F_1(a,b;c;z)$ is the hypergeometric function defined as
\begin{eqnarray}
    _2F_1(a,b;c;z)=\sum_{i=0}^{\infty}\frac{(a)_i (b)_i}{(c)_i}\frac{z^i}{i!}.
\end{eqnarray}
Here, $(x)_n$ denotes the Pochhammer symbol given by $(x)_n=\Gamma{(x+n)}/\Gamma{(x)}$. The quantity, $F_{SP}$, can be evaluated explicitly as a function of $N$ for a given $k$. After optimizing $F_{SP}$ with respect to $\theta$, one can obtain the maximum QFI in this scenario, which we denote by $F_{SP}^{max}$. Below, we evaluate $F_{SP}^{max}$ exactly for $k=2, 3$,
and find the scaling of $F_{SP}^{max}$ with $N$ for higher values of $k$ by 
numerical analysis in the limit of $N\gg k$. 
In this large $N$ limit, numerics dictate that
the maximal QFI vary as 
$ \beta N^{\alpha}$. The best-fit values of the parameters, $\alpha$ and $\beta$ corresponding to $k=4,5$ and $6$ are presented in Table~\ref{scale} with appropriate error bars. From Table~\ref{scale}, we find that for $k$-body interactions, and optimal symmetric probes, the value of $\alpha \approx 2k-1$. The exact expressions of $F_{SP}^{max}$ for $k=2$ and $3$ are provided in the two succeeding sub-subsections.
\begin{table*}[!htb]
\centering
\begin{tabular}{|c|c|c|c|}
\hline
   \; $k$ \; & \;\;\; $\widetilde{\alpha} \pm \delta$ \;\;\; &  \;\;\;\;\;\; $\widetilde{\beta} \pm \delta$ \;\;\;\;\;\; &  \;\;\;\;\;\; $\mathcal{R}$ \;\;\;\;\;\;   \\  
  \hline
  \hline
  4 & 7.00861 $\pm$ 0.000060211 & 0.010873427 $\pm$ 5.08395 $\times 10^{-6}$ & 0.000609672  \\ 
   \hline
   5 & 9.005658 $\pm$ 0.0000613882 & 0.0005266667 $\pm$ 2.51061 $\times 10^{-7}$  & 0.000621592  \\ 
   \hline
  6 & 11.01 $\pm$ 0.0000938575 & 0.0000165369 $\pm$ 1.20526 $\times 10^{-8}$  & 0.000950364  \\ 
   \hline
   \end{tabular}
\caption{Scaling of maximum QFI with the number of parties  for optimal symmetric product input probes in the limit $N \gg k$. All the estimated parameters are presented in a manner similar to that in Table~\ref{table:opt}, and they are obtained by following the least-squares method of Appendix~\ref{andhare_alo}. For instance, $\widetilde{\beta}$ and $\delta$ denote the maximum likelihood estimator and the error bar, respectively, for the fitting parameter, $\beta$. The convergence of the values of $\widetilde{\alpha}$ and $\widetilde{\beta}$ has been checked at  values of $N= 3000$.}
\label{scale}
\end{table*}
\subsubsection{Two-body interactions}
Here we consider the case of two-body interactions, i.e. $k=2$, and maximize the quantity, $F_{SP}$ (refer to Eq.~\eqref{sp}), with respect to $\theta$. The QFI in this case is given by
\begin{eqnarray}
    F_{SP}^{(k=2)}=\beta_0+\beta_1 z + \beta_2 z^2,  
\end{eqnarray}
where the coefficients are $\beta_i=\binom{N}{i}\binom{N-i}{i}\binom{N-2i}{2-i}$ for $i=0,1,2$, and $z=\cos^2\theta$. The quantity, $F_{SP}^{(k=2)}$, is optimal
corresponding to two values of $\theta$, say $\theta_0$ and $\theta_1$, where one of them, say
$\theta_1$, satisfies the 
equation,
\begin{equation}
    \theta_1 = \cos^{-1} \left(\sqrt{-\frac{\beta_1}{2 \beta_2} }\right).
\end{equation}
For $N>2$, the quantity, $-\frac{\beta_1}{2 \beta_2}>0$, and  $\theta_1$ corresponds to the maxima of $F_{SP}^{(k=2)}$. Therefore $F_{SP}^{max(k=2)}$ for $N>2$ is given by
\begin{equation}
    F_{SP}^{max(k=2)} = 4 \Delta^2 h_2^{(N)} = \frac{2N(N-1)^3}{(2N-3)}.
    \label{n2}
\end{equation}
On the other hand, $\theta_0=(2n+1)\pi/2$, where $n$ is an integer, gives the maximum $F_{SP}$ for $N=2$, and the corresponding maximal value is given by $F_{SP(N=2)}^{max(k=2)}=4$. From Eq.~\eqref{n2}, we can conclude that in the limit of large $N$, 
the maximum QFI for two-body interacting encoding Hamiltonians scales as $F_{SP}^{max(k=2)}\sim N^3$. 

\subsubsection{Three-body interactions}
We repeat the same procedure for calculating $F_{SP}^{max}$ in the case when $k=3$.
In this case, the QFI, as follows from Eq.~\eqref{sp}, is given by
\begin{eqnarray}
    F_{SP}^{(k=3)}=\gamma_0+\gamma_1 z + \gamma_2 z^2 +\gamma_3 z^3,  
\end{eqnarray}
where the coefficients are as follows: $\gamma_i=\binom{N}{i}\binom{N-i}{i}\binom{N-2i}{3-i}$, for $i=0$ to $3$ and $z=\cos^2\theta$. 
After optimizing $F_{SP}^{(k=3)}$ with respect to $\theta$, we obtain two values of $\theta$  belonging to the set, $\{\widetilde{\theta}_0,\widetilde{\theta}_1 \}$, which give the maxima 
for different $N$. The first one is $\widetilde{\theta}_0=(2n+1)\pi/2$, which gives the maximum for $N=3$, and the corresponding maximum QFI is $F_{SP(N=3)}^{max(k=3)}=4$. The other solution, $\widetilde{\theta}_1$  produces the maximum for 
$N>3$, and satisfies the following equation:
\begin{equation}
\cos^2\widetilde{\theta}_1=\frac{-\gamma_2\pm \sqrt{\gamma_2^2-3\gamma_1\gamma_3}}{3\gamma_3}.
\end{equation}
The closed form of $F_{SP}^{max}$ for three-body interaction is given in terms of $N$ as
 \begin{widetext}
 \begin{eqnarray}
     F=\frac{2 (N-2) N \left((\text{N1}-3) ((N-3) N+4) x+(N-1) \left(\text{N1} (N-3)^2+4\right) (N-2)^3\right)}{3 (3 (N-5) N+20)^2},
 \end{eqnarray}
 \end{widetext}
 where $x=\sqrt{(N-3) (N-2)^3 (N-1)^2 ((N-3) N+4)}$.
From the exact analytical form of the QFI, we can infer that for three-body interacting encoding Hamiltonians, the maximal QFI obtainable using symmetric product input probes varies as $\sim 4N^5/27$, in the limit $N\gg k$.

\subsection{Remarks}
In the preceding subsections, we have considered two scenarios under which we find the maximum attainable precision in estimating the coupling strength of the encoding Hamiltonian, viz. encoding onto optimal symmetric product probes and (general, i.e., unrestrained) optimal probes. The measurement strategy considered is optimal. We find that the scaling of  maximum QFI with the number of system particles in each of these scenarios monotonically increases with increasing value of $k$, where $k$ denotes $k$-body interactions. In particular, if the maximum QFI varies as $\sim \beta N^{\alpha}$  in the limit of  number of system particles, $N$, much larger in comparison to $k$, then for symmetric product probes, $\alpha$ is $2k-1$, whereas for optimal probes, $\alpha$ is $2k$. Moreover, it is interesting to note that the scaling of $N$ is exactly one order less in symmetric product inputs than in the optimal ones. So the values of $\alpha$ gradually increase in order as $3,4,5,6, \ldots$ and so on, as the scenarios switch from symmetric product two-body case to optimal two-body case to symmetric product three-body case to optimal three body case to symmetric product four-body case and so on. The coefficient, $\beta$, however reduces with increasing $k$, although at a rate slower than that of increase of $\alpha$, for large number of particles. It is further to be noted that no apparent dichotomy  is observed for even- and odd-body interactions in the scaling of the maximum QFI in both the scenarios, viz. product and general probes.

\section{Entanglement vs. asymmetry in optimal probes for even- and odd-body interactions}
\label{mayabi}
In this section, we find some intriguing features of the optimal probes, separately for even- and odd-body interacting encoding Hamiltonians, in the estimation of coupling constant. Let us first consider the even-body case. 


\subsection{Even-body interactions}
\label{amaltas}
When $k$ is even, one of the eigenvectors of the Hamiltonian, $h_k^{(N)}$, 
is given by $\ket{E_{m}}=\ket{0}^{\otimes m}\otimes \ket{1}^{\otimes N-m}$. 
The optimal value of $m$, i.e. $m_0$ which minimizes the eigenvalue, $E_m$, corresponds to the eigenstate $\ket{E_{m_0}}$.
The situation $m=0$, would imply that 
the eigenvector does not have any spin pointing upwards (i.e. $\ket{0}$), and therefore the corresponding eigenvector is $\ket{1}^{\otimes N}$, and since  $k$ is even, the corresponding minimum eigenvalue is $\binom{N}{k}$.
This would in turn suggest that the Hamiltonian is fully degenerate, which is clearly a contradiction, since the Hamiltonian consists of tensor products of Pauli-$z$. 
This proves that the eigenstate corresponding to the minimum eigenvalue of any even-body encoding Hamiltonian considered here is never of the form $\ket{1}^{\otimes N}$. 
Further, the eigenvector corresponding to the maximum eigenvalue for $k$-body interactions, where $k$ is even, is given by $\ket{E_{max}}=\ket{0}^{\otimes N}$, with the corresponding eigenvalue being $E_{max}=\binom{N}{k}$. 
Since $m\ne 0$,  the optimum eigenstate is never genuine multiparty entangled.

Let us here discuss the case of $k=2$ in detail.
The discussion in subsection~\ref{even} leads us to the following inference. 
If the number of particles is odd, then a possible choice of eigenstate corresponding to the minimum  eigenvalue of $h_2^{(N)}$ is given by $\ket{E_m^{o}}=\ket{0}^{\otimes (N-1)/2}\otimes\ket{1}^{\otimes (N+1)/2}$ or $\ket{E_m^{o}}=\ket{1}^{\otimes (N-1)/2}\otimes\ket{0}^{\otimes (N+1)/2}$, while if $N$ is even, then one choice of minimum eigenstate is $\ket{E_m^{e}}=\ket{0}^{\otimes N/2}\otimes\ket{1}^{\otimes N/2}$. (The superscripts $o$ and $e$ indicates whether the state corresponds to odd or even values of $N$ respectively.) Whereas, the maximum-energy eigenstate of $h_2^{(N)}$ has all  spins either pointing  upwards or downwards in the \(z\)-direction. Without loss of generality, let us consider the maximum energy eigenstate to be the one with all spin up i.e. $\ket{E_{M}^{o/e}}=$ $\ket{1}^{\otimes N}$. We also consider that the minimum eigenstate corresponding to odd values of $N$ have  $(N-1)/2$ up spins and $(N+1)/2$ down spins, i.e. $\ket{E_{M}^{o}}=$ $\ket{0}^{\otimes (N-1)/2}\otimes \ket{1}^{\otimes (N+1)/2}$. The resulting optimum state for even and odd values of $N$ thus becomes
\begin{eqnarray}
\label{opt_st}
     \ket{\chi_e}=\frac{1}{\sqrt{2}}\left(\ket{0}^{\otimes N/2}+\ket{1}^{\otimes N/2}\right)\otimes \ket{1}^{\otimes N/2} , \;\;\; \text{and} \nonumber \\
      \ket{\chi_o}=\frac{1}{\sqrt{2}}\left(\ket{0}^{\otimes (N-1)/2}+\ket{1}^{\otimes (N-1)/2}\right)\otimes  \ket{1}^{\otimes (N+1)/2} \;\;\;\;
\end{eqnarray}
respectively.
Interestingly we find that both $\ket{\chi_e}$ and $\ket{\chi_o}$ are asymmetric, in a sense that all $l$-party reduced states are not the same for $l=1$ to $N$. This feature is in contrary to the case of $h_1^{(N)}$, where the optimum state is a genuine $N$-party entangled state. In fact for $N\le 3$, it is striking to note that the optimum input state may be asymmetric product. For $N=2$, such an optimum probe is given by $\ket{\chi_e}=\ket{+1}$, where $\ket{+}=(\ket{0}+\ket{1})/\sqrt{2}$. Likewise for $N=3$, the state is $\ket{\chi_o}=\ket{+11}$. Thus on this account one can claim that for two-body generators, for values of $N\ge2$, asymmetry in input probes is a bonafide resource in parameter estimation protocols.

\begin{figure}
\centering
\includegraphics[width=8cm]{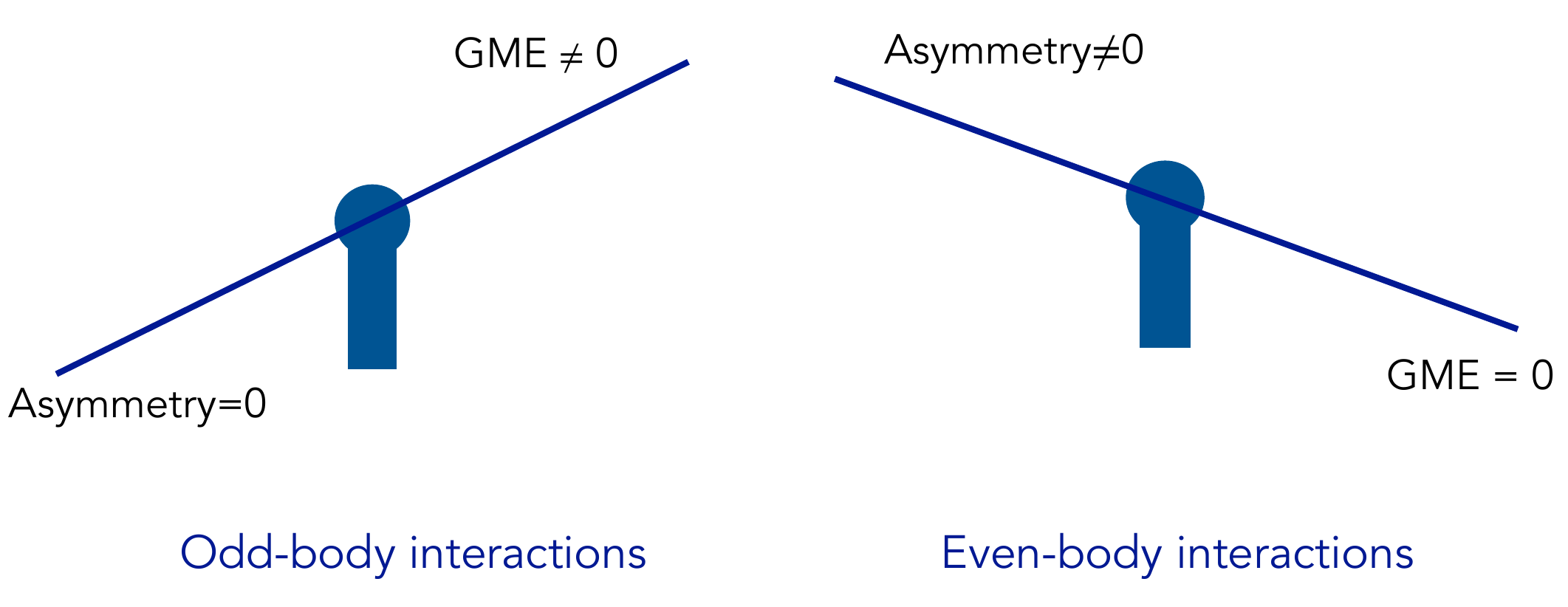}%
\caption{
Complementarity vis-{\`a}-vis resource requirement in optimal probes for estimating strength of odd- and even-body interactions.
For odd-body interactions, genuine multiparty entanglement is necessary in the optimal probes, but asymmetry in such states is not crucial. However for even-body interactions, asymmetric probes are useful in attaining optimality, though genuine multipartite entanglement is not necessary in the optimal probes in this case.}
\label{fig_ss}
\end{figure}

\subsection{Remarks}
\begin{figure}
\centering
\includegraphics[width=8cm]{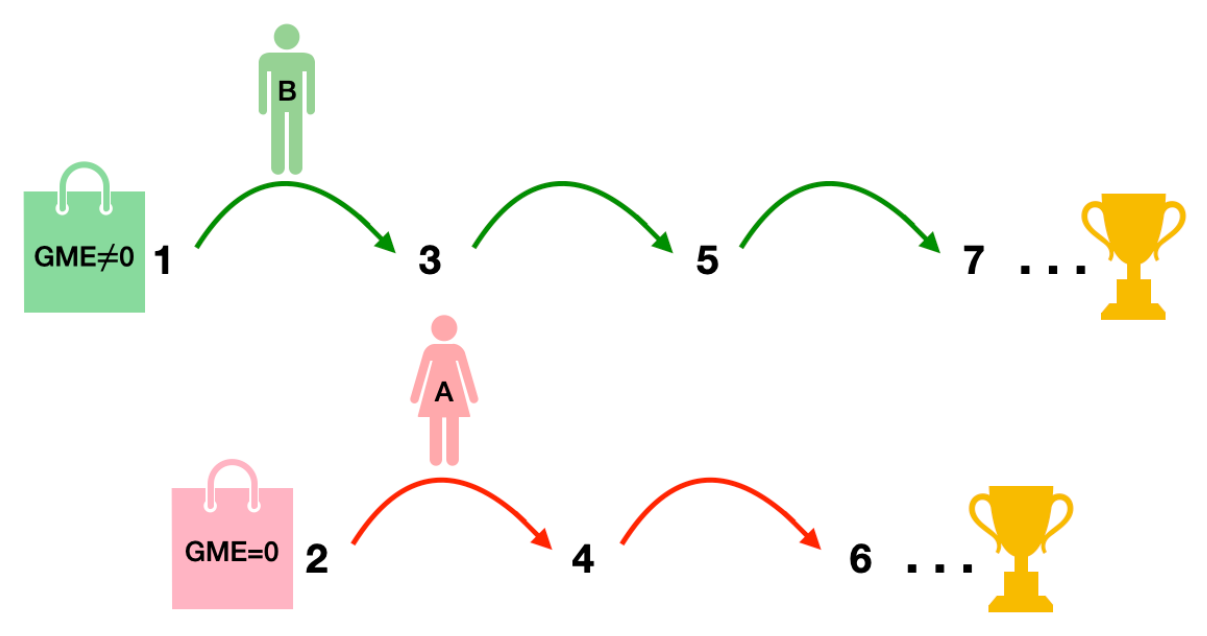}%
\caption{The schematic depicts  dichotomy between the presence and absence of genuine multipartite entanglement in the optimal input probes for $k$-body interactions with odd and even values of $k$ respectively. If Alice (A) encounters only the values of $k$ which are even, denoted by the red arrows, then she does not require any genuine multipartite entangled input state to achieve the best metrological precision under the relevant settings. However, if Bob (B) comes across odd values of $k$, denoted by the green arrows, then he would inevitably require genuine multipartite entangled probes for the best precision.}
\label{fig}
\end{figure}

The results in this section infer that while odd-body interacting encoding Hmailtonians necessitate genuinely multiparty entangled probes, even-body interactions require asymmetry instead of genuine multiparty entanglement in them. Fig.~1 of main text schematically depicts this feature for even- and odd-body encoding Hamiltonians. Another interesting feature is the dichotomy between the presence and absence of genuine multiparty entanglement in optimal probes for odd- and even-body interactions respectively. This feature is depicted in Fig.~\ref{fig}.

So far we have only considered the local component of the generator to be $\sigma_{z}$ with eigenvalues $+1$ and $-1$. In the next section, we generalise the case to a generator of arbitrary local dimension, and establish certain conditions on the  generator that ensure asymmetry in optimal probe state.

\section{Higher-dimensional probes}
\label{higher}
Here we present the analyses of estimation of coupling strengths for two- and three-body encoding Hamiltonians, where the dimension of each subsystem is arbitrary. The Hamiltonians correponding to two- and three-body interactions considered are respectively given by
\begin{eqnarray}
     \widetilde{h}_2^{(3)}=J\sum_{\substack{i=1\\j>i}}^3 H_i H_{j} \;\;\; \text{and} \;\;\;  \widetilde{h}_3^{(4)}=J\sum_{\substack{i=1\\ k>j>i}}^4 H_i H_{j}H_k
\end{eqnarray}
The notation, $H_j$, here indicates that the operator, $H$, acts locally on the $j^{th}$ party.
For two-body interactions in arbitrary dimensions, we consider $3$ parties, whereas for three-body interactions, we consider $4$ parties. As discussed earlier, only the maximum and minimum eigenvalues of the total Hamiltonian is required to calculate the QFI and this statement is true irrespective of the local dimension of the probe parties~\cite{variance}. Thus in order to find the optimal state that maximizes the QFI for a Hamiltonian of arbitrary local dimension, one needs to first find the maximum and minimum eigenvalues of the local component of the Hamiltonians. To do so, we consider the maximum and minimum eigenvalues of the local component of the Hamiltonian, $H$, to be equal to $\delta_M$ and $\delta_m$ respectively with the relevant eigenvectors $\ket{\delta_M}$ and $\ket{\delta_m}$. 
Now there can arise four scenarios, which are given by
\begin{itemize}
    \item $\mathcal{A}_1$: $\delta_M >0 $ and $\delta_m >0 $
     \item $\mathcal{A}_2$: $\delta_M < 0 $ and $\delta_m < 0 $
      \item $\mathcal{A}_3$: $\delta_M >0 $, $\delta_m <0 $ and $|\delta_M|>|\delta_m|$
       \item $\mathcal{A}_4$: $\delta_M >0 $, $\delta_m < 0 $ and $|\delta_M|<|\delta_m|$
         \item $\mathcal{A}_5$: $\delta_M >0 $, $\delta_m < 0 $ and $|\delta_M|=|\delta_m|$.
\end{itemize}
We analyze these five scenarios, first for Hamiltonians with two-body interactions and then for three-body interactions on the basis of the fact that one can fully characterize the maximal QFI and the relevant optimal state only in terms of the maximum and minimum eigenvalues of $H$.
\subsection{Two-body interactions}
 Let us first consider a two-body generator, $\widetilde{h}_2^{(3)}$, comprising of three parties. Our goal is to maximize the variance of $\widetilde{h}_2^{(3)}$, and find the corresponding optimal state. 
A possible set of eigenvectors and eigenvalues of $\widetilde{h}_2^{(3)}$ is given by
\begin{eqnarray}
    (a) &:& \ket{D_1} = \ket{\delta_M}^{\otimes 3} \rightarrow 3\delta_M^2 \nonumber \\
    (b) &:&   \ket{D_2} = \ket{\delta_m}^{\otimes 3} \rightarrow 3\delta_m^2 \nonumber \\
    (c) &:& \ket{D_3} = \ket{\delta_m}\otimes\ket{\delta_M}^{\otimes 2}  \rightarrow 2\delta_m\delta_M + \delta_M^2 \nonumber \\
    (d) &:& \ket{D_4} = \ket{\delta_M}\otimes\ket{\delta_m}^{\otimes 2}  \rightarrow 2\delta_m\delta_M + \delta_m^2. \nonumber
\end{eqnarray}
  The eigenvalues in the respective cases is denoted by the indices $a, b, c$ and $d$. Note that each of the eigenvectors are symmetric under permutation with respect to different parties. The maximum and minimum eigenvalues of the total generator belong to the set $\{a,b,c,d\}$. It, however, differs for different settings, i.e. $\mathcal{A}_i$, for $i=1$ to $5$. We analyze the different settings and find the maximum and minimum eigenvalues corresponding to each setting in the following discussion.
\\
\\
\textbf{$\mathcal{A}_1$:} 
Since $a>b,c,d$ in this case, the maximum eigenvalue of $h_2^{(3)}$ is $a=3\delta_M^2$ with the corresponding eigenvector $\ket{D_1}=\ket{\delta_M}^{\otimes 3}$. 
Further, as $b<a,c,d$, we can say that $b=3\delta_m^2$ is the minimum eigenvalue with eigenvector $\ket{D_2}=\ket{\delta_m}^{\otimes 3}$. Thus one can conclude that under the setting $\mathcal{A}_1$, the optimum state that maximises the QFI is given by $\ket{A_1}=\left(\ket{\delta_M}^{\otimes 3}+\ket{\delta_m}^{\otimes 3}\right)/\sqrt{2}$, which is a GHZ state.
\\
\\
\textbf{$\mathcal{A}_2$:} 
In this scenario, since $a<b,c,d$, the minimum eigenvalue is $a=3\delta_M^2$ corresponding to the eigenvector $\ket{\delta_M}^{\otimes 3}$. Similarly, we find that $b>a,c,d$, and thus $b=3\delta_m^2$ is the maximum eigenvalue in this case with eigenvector $\ket{\delta_m}^{\otimes 3}$. So, it's straight forward to infer that the optimum eigenstate, $\ket{A_2}$, that maximises QFI is same as the previous case, i.e., $\ket{A_2}=\ket{A_1}$. Note that this state has non-zero GME.
\\
\\
\textbf{$\mathcal{A}_3$:} In this case, 
$a>b,c,d$, and the maximum eigenvalue here is $a=3\delta_M^2$ with the corresponding eigenvector $\ket{\delta_M}^{\otimes 3}$. 
The minimum eigenvalue in the second case is $d=2\delta_m\delta_M + \delta_m^2$ corresponding to the eigenvector, $\ket{D_3}=\ket{\delta_M}\otimes\ket{\delta_m}^{\otimes 2}$.
Therefor the optimal state in this scenario which maximises the QFI is $\ket{A_3}=\ket{\delta_M}\left(\ket{\delta_M}^{\otimes 2}+\ket{\delta_m}^{\otimes 2}\right)/\sqrt{2}$, which is clearly not genuine multiparty entangled.
\\
\\
\textbf{$\mathcal{A}_4$:}
In this setting, $c<d<a<b$. Therefore the minimum and maximum eigenvalues are respectively $c=2\delta_m\delta_M + \delta_M^2$ and $b=3\delta_m^2$ with their relevant eigenvectors. So the state which gives the maximal QFI in this case is $\ket{A_4}=\ket{\delta_m}\left(\ket{\delta_m}^{\otimes 2}+\ket{\delta_M}^{\otimes 2}\right)/\sqrt{2}$, which again has GME equal to zero.
\\
\\
\textbf{$\mathcal{A}_5$:} There can be one last situation corresponding to $\mathcal{A}_5$, i.e. the case where $|\delta_m|=|\delta_M|=\delta$. The maximum eigenvalue of $h_2^{(3)}$ is $d_M=3\delta^2$, and the minimum eigenvalue is $d_m=2\delta_m\delta_M+\delta_m^2=2\delta_M\delta_m+\delta_M^2=-\delta^2$.
The maximum eigenvalue $d_M$ corresponds either to the state $\ket{D_1}$ or $\ket{D_2}$. Similarly, the minimum eigenvalue will correspond to either the state, $\ket{D_4}$ or $\ket{D_3}$. States that maximize the variance can have the following choices 
\begin{eqnarray}
    \ket{O_1} &=& \ket{\delta_M}\left(\ket{\delta_M}^{\otimes 2}+\ket{\delta_m}^{\otimes 2}\right)/\sqrt{2}, \nonumber \\
    \ket{O_2} &=& \ket{\widetilde{+}\delta_M\delta_M}, \nonumber \\
    \ket{O_3} &=& \ket{\delta_m}\left(\ket{\delta_m}^{\otimes 2}+\ket{\delta_M}^{\otimes 2}\right)/\sqrt{2}, \nonumber \\
    \ket{O_4} &=& \ket{\widetilde{+}\delta_m\delta_m}, \nonumber
\end{eqnarray}
where $\ket{\widetilde{+}}=\left(\ket{\delta_M}+\ket{\delta_m}\right)/\sqrt{2}$. 
Note that all these choices of states have GME $=$ 0. Moreover, the states $\ket{O_2}$ and $\ket{O_4}$, are asymmetric product states.
Thus the scenario $\mathcal{A}_5$, exactly matches our observation in ~\ref{amaltas}, with $|\delta_M|=|\delta_m|=1$. 
\\
\\
It can be thus concluded that irrespective of the dimension of the generator,  asymmetric input probes can be utilised to attain the maximum QFI, provided that the minimum and maximum eigenvalues of the local component of the generator are of different signs.
However if both $\delta_m$ and $\delta_M$ are of same sign, then states having GME  are necessary to attain the maximum QFI.
A detailed discussion of the maximum and minimum eigenvalues of $\widetilde{h}_2^{(3)}$ corresponding to each of the five settings is provided in Section~\ref{appen:1}.
\subsection{Three-body interactions}
 In this subsection we consider the three-body encoding Hamiltonian, $\widetilde{h}_3^{(4)}$ to see whether asymmetry still helps as a resource in estimating  the coupling constant. The Hamiltonian corresponding to the local subsystem is $H$ as before, having minimum and maximum eigenvalues $\delta_m$ and $\delta_M$ corresponding to eigenvectors $\ket{\delta_m}$ and $\ket{\delta_M}$, respectively. 

A set of five possible eigenvalues and corresponding eigenvectors of $h_3^{(4)}$ is given by
\begin{eqnarray}
    (a_0) &:& \ket{\delta_M}^{\otimes 4} \rightarrow 4\delta_M^3 \nonumber \\
    (b_0) &:& \ket{\delta_m}^{\otimes 4} \rightarrow 4\delta_m^3 \nonumber \\
    (c_0) &:& \ket{\delta_m}\otimes\ket{\delta_M}^{\otimes 3}  \rightarrow 3\delta_m\delta_M^2 + \delta_M^3 \nonumber \\
    (d_0) &:& \ket{\delta_M}\otimes\ket{\delta_m}^{\otimes 3}  \rightarrow 3\delta_m^2\delta_M + \delta_m^3 \nonumber \\
    (e_0) &:& \ket{\delta_M}^{\otimes 2}\otimes\ket{\delta_m}^{\otimes 2}  \rightarrow 2\delta_m\delta_M \left(\delta_m+\delta_M\right). \nonumber
\end{eqnarray}
We can again contemplate the predescribed five scenarios, $\mathcal{A}_i$, for $i=1$ to $5$, and find out the maximum and minimum eigenvalues of $h_3^{(4)}$ in a manner similar to that in the previous subsection.
\\
\\
\textbf{$\mathcal{A}_1$:} In this case, the minimum and maximum eigenvalues are given respectively by $b_0=4\delta_m^3$ and $a_0=4\delta_M^3$. Therefore in this case the optimal state that maximises the QFI is a GHZ state of the form $\ket{\widetilde{A}_1}=\left(\ket{\delta_M}^{\otimes 4}+\ket{\delta_m}^{\otimes 4}\right)/\sqrt{2}$.
\\
\\
\textbf{$\mathcal{A}_2$:} Arguing similarly, it can be shown that the minimum and maximum eigenvalues are $a_0=4\delta_M^3$ and $b_0=4\delta_m^3$ respectively. Therefore the optimal state in this case which gives the maximum QFI is again a GHZ state of the form $\ket{\widetilde{A}_2}=\ket{\widetilde{A}_1}$.
\\
\\
\textbf{$\mathcal{A}_3$:} The maximum eigenvalue of $h_3^{(4)}$ in this case is given by $a_0=4\delta_M^3$. 
Also in this scenario, $b_0<a_0,c_0,d_0$. 
Now let us consider the term $b_0-e_0=|\delta_m|\left(|\delta_m|+|\delta_M|\right)\left(-2|\delta_m|+|\delta_M|\right)$. This quantity can be either negative or positive depending upon whether $|\delta_M|<2|\delta_m|$ or $|\delta_M|>2|\delta_m|$, respectively. From the first condition, we can infer that $b_0<a_0,c_0,d_0,e_0$, and therefore $b_0$ is the minimum eigenvalue. In this case, the optimal input state is a GHZ of the form $\ket{\widetilde{A}_3^i}=\ket{\widetilde{A}_1}$. Under the second condition, i.e. $|\delta_M|>2|\delta_m|$, the ordering of the eigenvalues is given by $e_0<b_0<c_0,d_0<a_0$. In this case, the minimum eigenvalue is $e_0=2\delta_m\delta_M \left(\delta_m+\delta_M\right)$, and optimum state which gives the maximum QFI has the form $\ket{\widetilde{A}_3^{ii}}=\ket{\delta_M}^{\otimes 2}\left(\ket{\delta_M}^{\otimes 2}+\ket{\delta_m}^{\otimes 2}\right)/\sqrt{2}$, which possess zero GME.
\\
\\
\textbf{$\mathcal{A}_4$:} It can be argued similarly that here the minimum eigenvalue is always $b_0=4\delta_m^3$. Whereas the maximum eigenvalue depends upon two conditions, i.e. $|\delta_M|>|\delta_m|/2$ and $|\delta_M|<|\delta_m|/2$. In the former case, the maximum eigenvalue is $a_0=4\delta_M^3$, while in the latter case, it is given by $e_0=2\delta_m\delta_M \left(\delta_m+\delta_M\right)$. Therefore in the former situation, the optimal state is again a GHZ of the form $\ket{\widetilde{A}_4^i}=\ket{\widetilde{A}_1}$, and in the latter situation, it is given by $\ket{\widetilde{A}_4^{ii}}=\ket{\delta_m}^{\otimes 2}\left(\ket{\delta_M}^{\otimes 2}+\ket{\delta_m}^{\otimes 2}\right)/\sqrt{2}$, which clearly possess zero GME.
\\
\\
\textbf{$\mathcal{A}_5$:} There can be a last scenario corresponding to $\mathcal{A}_5$, i.e. the case when $|\delta_m|=|\delta_M|=\delta$. Following a similar argument, it can be shown that the maximum eigenvalue of $h_2^{(3)}$ in this case is $a_0=4\delta_M^3$, and the minimum eigenvalue is $b_0=4\delta_m^3$. Thus the optimal input probe in this situation will always be a GHZ state given by $\ket{\widetilde{A}_5}=\ket{\widetilde{A}_1}$, which has non-zero GME.
\\
\\
We can conclude from the discussion in this section that except the two scenarios, $\mathcal{A}_3$ and $\mathcal{A}_4$, the optimum input state which maximizes the relevant QFI is always a GHZ state having non-zero GME. In the two cases, $\mathcal{A}_3$ and $\mathcal{A}_4$, optimum input probes with zero GME is also a possibility. The results in the main text correspond to the case when $\delta_M=1$ and $\delta_m=-1$, i.e. situation $\mathcal{A}_5$, and again in this case, one can argue that whatever be the dimension of local component of the generator, asymmetry will not help in parameter estimation, if we consider three-body interactions  between four parties.

\section{Least-Squares method}
\label{andhare_alo}
We numerically obtain the relevant maximal QFI, which depends on the number of parties, $N$. So we have a list of values $y_i$, where each $i$ corresponds to each $N$, and $y_i$ corresponds to the respective maximal QFI corresponding to $N$. To find a function $f(x_i,\alpha,\beta)$ that best fits the data $\{y_i\}$, we adopt the least-squares method~\cite{lsqft1,lsqft2}. Since QFI scales as $\sim \alpha N^{\beta}$, we fit the data with a function, $f(x_i,\alpha,\beta)=\alpha \ln x_i +\beta$. We denote the difference between numerically obtained value and the functional value as $r_i$, and define a quantity, $\chi^2$:
\begin{eqnarray}
    \chi^2=\sum_i \frac{r_i^2}{\sigma_i^2}=\sum_i \frac{\left[y_i-f(x_i,\alpha,\beta)\right]^2}{\sigma_i^2}.
\end{eqnarray}
In our calculations, we consider all the standard deviations, $\sigma_i$, to be equal, since the data points $y_i$ themselves do not have any error bars. Minimizing $\chi^2$ over the set of real parameters $\{\alpha, \beta\}$, we obtain the least-squares (maximum likelihood) estimate, $\{\widetilde{\alpha}, \widetilde{\beta}\}$, and the function $f(x_i,\widetilde{\alpha},\widetilde{\beta})$ that provides the best fit for a given data set. The best-fit function gives a minimum error, $\mathcal{R}$:
\begin{eqnarray}
\label{batas}
    \mathcal{R}=\sqrt{\frac{1}{\Omega-\Omega'}\left[y_i-f(x_i,\widetilde{\alpha},\widetilde{\beta})\right]^2}.
\end{eqnarray}
Next we calculate the confidence intervals for the two parameters. We
first compute the $\Omega \times \Omega'$ matrix, $W$, given by
\begin{eqnarray}
    W=\left( {\begin{array}{cc}
        \vdots & \vdots \\
        \frac{\partial f(x_i,\alpha,\beta)}{\partial \alpha} & \frac{\partial f(x_i,\alpha,\beta)}{\partial \beta} \\
        \vdots & \vdots
        \end{array} } \right).
\end{eqnarray}
Since the function to be fitted is linear, the standard errors, $S(.)$, of the parameters are given by
\begin{eqnarray}
S(\alpha)&=&\mathcal{R}\sqrt{\left[(W^{T}W)^{-1}\right]_{11}}, \;\; \text{and} \nonumber \\
S(\beta)&=&\mathcal{R}\sqrt{\left[(W^{T}W)^{-1}\right]_{22}}.
\end{eqnarray}
Finally, the $1-\nu$ marginal confidence interval for the parameter, $\alpha$, is given by $\widetilde{\alpha} \pm \delta$, where
\begin{eqnarray}
     \delta = S(\alpha) t\left(\Omega-\Omega', 1-\frac{\nu}{2} \right).
\end{eqnarray}
Here $t\left(\Omega-\Omega', 1-\nu/2 \right)$ is the $1-\nu/2$ percentile of the Student’s $t$ distribution with $\Omega-\Omega'$ degrees of freedom. Similarly, we can obtain the confidence interval for the parameter, $\beta$. In the two tables given in the paper, we provide the values of estimated parameters as per $\widetilde{\alpha/\beta} \pm \delta$ with the $95\%$ confidence level (that is, with $\alpha = 0.05$) and the corresponding error, $\mathcal{R}$ defined in~\ref{batas}.


\section{Maximum and minimum eigenvalues of two-body generator in arbitrary dimensions}
\label{appen:1}
\textbf{$\mathcal{A}_1$:} Since $\delta_M >\delta_m$, the difference of the eigenvalues $a-b=3(\delta_M^2 - \delta_m^2) >0$, which implies $a>b$. Also $a-c=2\delta_M(\delta_M-\delta_m)>0$, as $\delta_M>0$. Therefore we get $a>c$. Further, the difference $a-d$ can be written as $a-d=(\delta_M-\delta_m)(3\delta_M+\delta_m)>0$, since both $\delta_m$ and $\delta_M$ are positive. This suggests $a>d$. Therefore since $a>b,c,d$, the maximum eigenvalue of $h_2^{(3)}$ is $a=3\delta_M^2$ in this case with the corresponding eigenvector $\ket{D_1}=\ket{\delta_M}^{\otimes 3}$. 
Now let us find the minimum eigenvalue. We already found that $b<a$. Now the difference $b-d=2\delta_m(\delta_m-\delta_M)$ which is negative since $\delta_m < \delta_M$. This implies $b<d$. In a similar way $b-c=(\delta_m-\delta_M)(3\delta_m+\delta_M)$. Since $\delta_m < \delta_M$ and $\delta_m$ and $\delta_M$ are both positive, $b-c<0$, which implies $b<c$. Therefore as $b<a,c,d$, we can say that $b=3\delta_m^2$ is the minimum eigenvalue with eigenvector $\ket{D_2}=\ket{\delta_m}^{\otimes 3}$. 
\\
\\
\textbf{$\mathcal{A}_2$:} In this scenario, $|\delta_M|<|\delta_m|$, and therefore $a-c=2|\delta_M| \left(|\delta_M|-|\delta_m|\right)<0$. Again, $a-d=\left(|\delta_M|-|\delta_m|\right)\left(3|\delta_M|+|\delta_m|\right)<0$, which implies $a<c,d$. Also in this case, $a<b$. Therefore since $a<b,c,d$, the minimum eigenvalue is $a=3\delta_M^2$ in this case corresponding to the eigenvector $\ket{\delta_M}^{\otimes 3}$. Following a similar argument, we find that $b>a,c,d$, and thus $b=3\delta_m^2$ is the maximum eigenvalue in this case with eigenvector $\ket{\delta_m}^{\otimes 3}$. 
\\
\\
\textbf{$\mathcal{A}_3$:} In this case, 
$a>b,c,d$, and therefore the maximum eigenvalue here is $a=3\delta_M^2$ with the corresponding eigenvector $\ket{\delta_M}^{\otimes 3}$. 
To find the minimum eigenvalue, we observe that $b<a$ as in the previous cases. The difference $b-d=2|\delta_m|(|\delta_m|+|\delta_M|)$ is a positive quantity. However, to analyse the quantity $b-c=\left(|\delta_m|+|\delta_M|\right)\left(3|\delta_m|-|\delta_M|\right)$, two situations arise further. First is when the expression, $\left(3|\delta_m|-|\delta_M|\right)$ is negative, implying, $b<c$. Therefore under this condition, we get $b<a,c$ and $d<b$. and the minimum eigenvalue of the generator, $h_2^{(3)}$, in such case is $d=2\delta_m\delta_M + \delta_m^2$ with eigenvector, $\ket{D_4}=\ket{\delta_M}\otimes\ket{\delta_m}^{\otimes 2}$. Second case, arises when $\left(3|\delta_m|-|\delta_M|\right)$ is positive. This implies, $b<a$ and $b>c,d$. Therefore, we need to find the lowest among $c$ and $d$. To do so, we calculate $c-d=\delta_M^2-\delta_m^2>0$, this suggests that the sequence of ordering of the eigenvalues is $d<c<b<a$. Hence the minimum eigenvalue in the second case is also $d=2\delta_m\delta_M + \delta_m^2$ corresponding to the eigenvector, $\ket{D_3}=\ket{\delta_M}\otimes\ket{\delta_m}^{\otimes 2}$.
The optimal state in this scenario which maximises the QFI is $\ket{A_3}=\ket{\delta_M}\left(\ket{\delta_M}^{\otimes 2}+\ket{\delta_m}^{\otimes 2}\right)/\sqrt{2}$, which is clearly not genuine multiparty entangled.
\\
\\
\textbf{$\mathcal{A}_4$:} Arguing in a similar manner in this scenario, we obtain $a<b$ and $a-c=2|\delta_m|(|\delta_m|+|\delta_M|)>0$. Now let us consider the quantity, $a-d=\left(|\delta_m|+|\delta_M|\right)\left(3|\delta_M|-|\delta_m|\right)$. Since, $3|\delta_M|<|\delta_m|$ is contradicts the condition of setting $\mathcal{A}_4$, the only option is $|\delta_M|>|\delta_m|/3$, which, in turn, implies $a>d$. Therefore we find, $c,d<a<b$. Now, we just have to compare between $c$ and $d$ to get the minimum eigenvalue. The expression, $c-d$, is again negative, which finally gives, $c<d<a<b$. Therefore the minimum and maximum eigenvalues are respectively $c=2\delta_m\delta_M + \delta_M^2$ and $b=3\delta_m^2$ with their relevant eigenvectors.
\\
\\
\textbf{$\mathcal{A}_5$:} The last situation corresponds to $\mathcal{A}_5$, i.e. the case where $|\delta_m|=|\delta_M|=\delta$. The maximum eigenvalue of $h_2^{(3)}$ is $d_M=3\delta^2$, and the minimum eigenvalue is $d_m=2\delta_m\delta_M+\delta_m^2=2\delta_M\delta_m+\delta_M^2=-\delta^2$.
The maximum eigenvalue $d_M$ corresponds either to the state $\ket{D_1}$ or $\ket{D_2}$. Similarly, the minimum eigenvalue will correspond to either the state, $\ket{D_4}$ or $\ket{D_3}$.

\bibliography{metrology}

\end{document}